\documentclass{article}
\usepackage[utf8]{inputenc}
\usepackage{xcolor}
\usepackage[numbers]{natbib}
\usepackage{listings}

\usepackage{booktabs}
\usepackage{tabularx}
\usepackage{booktabs}
\usepackage{amsmath}
\usepackage[htt]{hyphenat}
\usepackage{svg}
\usepackage{hyperref}
\usepackage{cleveref}

\usepackage{subcaption}
\usepackage{graphicx}
\begin{document}
\begin{center}

%Title
\begin{Large}

Student Report \\
\vspace{0.5cm}
\ LensKit-Auto\

\end{Large}
\vspace{0.5cm}
\begin{large}

\ Max Breit\ \\
\ Anass Amezian El Idrissi\ \\
\ Rishikesh Giriraj Kulkarni\ \\
\ Luca Quade\ \\

\vspace{0.5cm}
University of Siegen \\
\vspace{0.1cm}
Chair: Intelligent Systems Group (ISG) \\
\vspace{0.1cm}
Supervisor: Prof. Dr.-Ing. Joeran Beel \\
\vspace{0.1cm}
Advisor: Tobias Vente \\
\vspace{0.2cm}
\ March 2026 \\

\end{large}

\end{center}

\begin{abstract}
    \noindent Recommender systems have a wide area of application, e.g. in fields like video streaming, social media, or digital marketplaces. But, for a recommender-system, finding the right algorithm with the right hyperparameters is a reoccurring challenge. There is no one-fits-all solution, since the performance of one algorithm can vary immensely on different data sets. Due to the challenges of finding the right algorithm and the broad use of recommender-systems, it is of interest to create an Automated Recommender System (\textbf{AutoRecSys}) that takes on the task of finding the right algorithm-hyperparameter-combination for a given data set. In this work, we present the enhancement of LensKit-Auto, a framework introduced by Vente et al., that solves exactly this task of finding a fitting algorithm-hyperparameter-combination. LensKit-Auto's biggest strength lies in its ease of use, where it operates as a black-box, into which the user can feed their data set and receive the information of which algorithm and hyperparameters work best on this data set. In this work, we bring LensKit-Auto up to date, so that it works with the new version of its underlying framework, LensKit. We also implement further functionalities, such as the Tree Parzen Estimator as an additional optimization method, the ability to reuse the found algorithm, updated documentation, and the ability to visualize the optimization process. We also adapt an existing meta-learning framework to generate a suitable meta-dataset for LensKit-Auto, which could enable the integration of meta-learning into LensKit-Auto in the future. The presented changes bring LensKit-Auto up to date and enhance its usability, so that even non-experts in the field can find the right algorithm for their use case.
\end{abstract}

\newpage

\tableofcontents

\newpage

\section{Introduction}
Recommender Systems play a central role in modern online platforms by helping users discover relevant content among large collections of items \cite{Discovery}. Recommender Systems are used widely in domains such as e-commerce, video streaming, rental, social media, and digital marketplaces \cite{WideUse}. By analyzing patterns in user-item interactions, recommender systems predict user preferences and provide personalized recommendations \cite{PersonalizedRecs}.

Evaluating, comparing, and selecting recommender system algorithms usually requires a lot of experimentation. Algorithms are often evaluated on a specific dataset using ranking-based metrics such as normalized discounted cumulative gain (\textbf{nDCG}), recall, or hit ratio. The performance of recommender algorithms is very sensitive to the choice of the hyperparameters \cite{10.1145/3604915.3609488}, dataset characteristics \cite{datasetCharecteristics} and evaluation protocols \cite{EvaluationProtocols}. Manually configuring and evaluating recommender systems often ends up being very time-consuming and also prone to errors \cite{EvaluationProtocols}. Exploring large hyperparameter spaces via manual experimentation is impractical and often yields suboptimal results \cite{HyperparameterSearch}.

Automated approaches for hyperparameter optimization and experimentation have become increasingly more popular in machine learning and recommender system research \cite{AutoMLPopularity}. Automated Machine Learning (AutoML) techniques aim to reduce manual effort by automatically searching for suitable hyperparameter configurations with predefined search spaces. Such approaches support systematic experimentation and improve reproducibility by providing structured optimization procedures. These approaches also make the field of machine learning accessible for non-experts in the field.

LensKit-Auto is a project that enables automated configuration and evaluation of recommender systems. It is built as a wrapper around the LensKit framework, leveraging its modular architecture for recommender system experimentation. LensKit-Auto allows users to define hyperparameter search spaces, run optimization algorithms, and evaluate model configurations.
 
Overtime changes in the LensKit ecosystems introduced compatibility challenges for LensKit-Auto. The release of newer LensKit versions for LensKit, introduced substantial modifications to underlying APIs and dependencies, As a result LensKit-Auto continued to rely on outdated versions of LensKit and required updates to support the current ecosystems. Furthermore, the original project also lacked several features that are important for modern experiment workflows, like quality automated testing, experiment visualization and a greater choice of optimization strategies.

The goal of this project was to modernize, extend, and improve the LensKit Auto Framework so that it can be used with recent versions of LensKit and Support improved experimentation workflows. Instead of developing a new framework from scratch, the project focused on updating the existing framework and extending its functionality.

Several improvements were implemented as part of this work. The codebase was updated to ensure compatibility with the current versions of LensKit and its dependencies. The testing infrastructure was revised by fixing existing tests and adding new tests which increased the overall test coverage and improved the reliability of the project. In addition, new functionality was introduced. A Tree Parzen Estimator (\textbf{TPE}) optimization algorithm was implemented as an additional hyperparameter optimization method. Model Persistence functionality was also implemented to allow trained models to be saved and reused for future experiments. Integration with DeepCave visualization tool was implemented to enable interactive analysis of optimization results. Finally, the project documentation was updated and expanded to improve usability for future users. 

With these improvements, LensKit Auto provides a more reliable and maintainable platform for automated recommender system experimentation.

\section{Related Work / Literature Survey}
Automated Recommender Systems (\textbf{AutoRecSys}) is closely related to the field of automated machine learning (\textbf{AutoML}), but while AutoML seems very diverse, AutoRecSys is just at its beginnings. In this section, we highlight some AutoRecSys tools and compare them with our tool, LensKit-Auto.

Auto-Surprise \cite{AutoSurprise} is a Python library that is, in the way it is used, pretty similar to LensKit-Auto. We can use data as input and let Auto-Surprise find the best algorithm and hyperparameter configuration for us. This algorithm is then returned with its tuned hyperparameters. Like the name suggests, Auto-Surprise is built upon the Surprise Library for Python, whereas we use LensKit. Another difference are the available algorithms. Auto-Surprise and LensKit-Auto share some algorithms, like SVD for example,  but both have algorithms that the other one does not have. Auto-Surprise has \textit{Slope One} for example, while LensKit-Auto can train a \textit{BiasedMFScorer}, a biased matrix factorization algorithm. Another minor detail is that, at the end of its run, Auto-Surprise outputs the performance of all its used algorithms as a dictionary, but LensKit-Auto prints this information only in the console during its run.

Another library is librec-auto  \cite{librecAuto}, which uses the librec library as its foundation. But other than LensKit-Auto, librec-auto lays its focus on finding the best hyperparamters for one algorithm. librec-auto also supports parallelism during its hyperparameter-optimization, which LensKit-Auto does not have, and also has an integration to messaging platforms like Slack, to notify the user when a run in librec-auto is finished. But all in all it seems like librec-auto's use-case differs a little from LensKit-Auto's, since librec-auto focuses more on the goal to make AutoRecSys experiments more reproducible.

\section{Background}
\subsection{Recommender Systems}

Recommender Systems rank items according to user preferences and interactions. Many online platforms contain extensive catalogs of products, videos, and music. Manually exploring such collections is difficult for users. Recommender systems use past interactions to rank items and present personalized suggestions. Applications include e-commerce, streaming services, social media systems, and digital libraries \cite{Discovery} 

Most recommender systems rely on historical interaction data between users and items \cite{Aggarwal2016Recommender}. This information is commonly represented as a user-item interaction matrix \cite{Aggarwal2016Recommender}. In a user-item interaction matrix, rows correspond to users and columns correspond to items. Each entry stores an observed interaction value, for example, a numerical rating or a binary value indicating whether a user interacted with an item \cite{Aggarwal2016Recommender}. Many entries remain unknown because each user interacts with only a small subset of available items \cite{Aggarwal2016Recommender}. As a result, the interaction matrix is typically sparse \cite{Aggarwal2016Recommender}. High sparsity makes it difficult to identify relevant items based on past interactions \cite{Aggarwal2016Recommender}. 

Collaborative filtering is one of the most successful approaches for generating recommendations \cite{Su2009CollaborativeFiltering}. Collaborative filtering exploits patterns within the interaction matrix to identify relationships between the users and items. The base assumption is that users who share similar preferences on a set of items are likely to prefer similar items with which they have not yet interacted \cite{Su2009CollaborativeFiltering}. Thus, the recommendations can be generated by identifying users or items with similar interaction patterns. 

Memory-based methods compute the similarity between relationships directly from the interaction matrix. User-based approaches identify users with similar preference patterns and recommend items with which similar users have interacted. Instead, Item-based approaches compute similarities between items and recommend items that resemble previously consumed content. Item-based collaborative filtering is widely applied in industrial recommender systems because it scales well to large datasets \cite{Sarwar2001ItemBasedCF}. 

In contrast to memory-based approaches, Matrix factorization is a widely used model-based method in recommender systems \cite{Su2009CollaborativeFiltering}. Matrix factorization represents a widely used model-based technique. In this approach, the interaction matrix is decomposed into two lower-dimensional matrices representing latent user and item factors. The latent factors capture underlying preference patterns that explain the observed interactions. Predictions are generated by computing the interaction between user and item factor vectors \cite{Koren2009MatrixFactorization}. Matrix factorization methods have demonstrated strong performance in many recommender system benchmarks and industrial applications. 

In addition to algorithmic approaches, recommender systems can also be categorized based on the type of feedback used for training. Recommender systems commonly operate on either explicit or implicit feedback. Explicit feedback refers to interactions in which users directly express preferences, for example by assigning a rating to items. This information provides a clear signal about user satisfaction, but often remains limited because users rarely provide ratings. 

Implicit feedback derives from user behavior, such as viewing, clicking, or purchasing items. Such interactions occur naturally during system usage and, therefore, produce larger datasets. Implicit signals do not directly represent preference strength. A missing interaction does not indicate dislike. Algorithms designed for implicit feedback must interpret interaction data differently and often focus on ranking relevant items rather than predicting rating values \cite{Hu2008Implicit}.

The evaluation objective depends on the recommendation task. Rating prediction tasks measure the accuracy with which a model predicts user ratings. Ranking-based tasks focus on identifying relevant items within a recommendation list. Ranking metrics such as normalized discounted cumulative gain (nDCG), recall or hit ratio measure how well relevant items appear within the top positions of the recommended lists \cite{Aggarwal2016Recommender}. Therefore, the choice of feedback type, model architecture, and evaluation protocol strongly influences the design and performance of recommender systems. 

\subsection{Evaluation of Recommender Systems}
The evaluation of recommender systems depends on both the recommendation task and the type of feedback available in the data \cite{herlocker2004evaluating, gunawardana2012evaluating, hu2008collaborative}. In general, recommender systems data can be divided into \emph{explicit feedback} and \emph{implicit feedback}. Explicit feedback consists of direct preference signals, such as numerical ratings or like/dislike statements. In contrast, implicit feedback is inferred from user behavior such as clicks, purchases, browsing activity, or watch events \cite{koren2009matrix, hu2008collaborative}. These two types of feedback lead to different evaluation settings.

\begin{table}[htp]
\centering
\caption{Example of an explicit-feedback dataset}
\label{tab:explicit-feedback-example}
\begin{tabular}{lccccc}
\toprule
\textbf{User} & \textbf{Movie A} & \textbf{Movie B} & \textbf{Movie C} & \textbf{Movie D} & \textbf{Movie E} \\
\midrule
User 1 & 5 & 4 & 1 & 5 & ? \\
User 2 & 4 & ? & 5 & 2 & 1 \\
User 3 & ? & 5 & 4 & 1 & 2 \\
\bottomrule
\end{tabular}
\end{table}

Table~\ref{tab:explicit-feedback-example} illustrates an explicit-feedback dataset in the form of a user-item rating matrix. Rows correspond to users, columns correspond to items, and each observed entry represents a rating directly provided by a user. Missing entries indicate user-item pairs for which no rating is available. In such settings, a common task is \emph{rating prediction}, where the goal is to estimate the rating value that a user would assign to an unseen item \cite{koren2009matrix, gunawardana2012evaluating}.

For example, if the rating of \emph{User 2} for \emph{Movie B} is hidden during evaluation, the recommender system predicts a value for this missing entry. The predicted value is then compared to the true rating. Error-based metrics such as Root Mean Squared Error (RMSE) and Mean Absolute Error (MAE) are commonly used because they measure how closely predicted ratings match the observed ratings \cite{gunawardana2012evaluating}. Lower RMSE and MAE values indicate more accurate predictions.

However, recommender systems are often not only required to predict rating values, but also to produce a ranked list of relevant items. This task is commonly referred to as \emph{Top-N recommendation} \cite{deshpande2004item, gunawardana2012evaluating}. In this setting, the objective is not to estimate the exact rating of an item, but to place the most relevant items at the top of the recommendation list \cite{gunawardana2012evaluating, herlocker2004evaluating}. Even when explicit ratings are available, they can be transformed into a ranking task by defining a relevance threshold, for example, by treating items with rating $\geq 4$ as relevant \cite{herlocker2004evaluating}.\\

\begin{table}[htp]
\centering
\caption{Example of an implicit-feedback dataset}
\label{tab:implicit-feedback-example}
\begin{tabular}{lccccc}
\toprule
\textbf{User} & \textbf{Movie A} & \textbf{Movie B} & \textbf{Movie C} & \textbf{Movie D} & \textbf{Movie E} \\
\midrule
User 1 & 1 & 1 & 0 & 1 & 0 \\
User 2 & 1 & 0 & 1 & 1 & 0 \\
User 3 & 0 & 1 & 1 & 0 & 1 \\
\bottomrule
\end{tabular}
\end{table}

Table~\ref{tab:implicit-feedback-example} shows a simple implicit-feedback dataset. Here, an entry of 1 indicates that an interaction was observed, for example a click, purchase, or watch event, while 0 indicates that no interaction was observed. In contrast to explicit feedback, implicit feedback does not directly express how much a user liked an item. Instead, it provides indirect evidence of possible interest \cite{koren2009matrix, hu2008collaborative}. In particular, the absence of an interaction does not necessarily mean that the user dislikes the item, it may simply mean that the item was never seen \cite{hu2008collaborative}.\\\\
Because implicit feedback does not provide exact preference scores, ranking-based evaluation is commonly used in this setting. The system generates a recommendation list and is then evaluated based on how well relevant items are placed in the top positions of the ranking. Metrics such as Precision@K, Recall@K, and normalized Discounted Cumulative Gain (nDCG@K) are commonly used for this purpose \cite{herlocker2004evaluating, gunawardana2012evaluating, jarvelin2002cumulated}.

\begin{table}[htp]
\centering
\caption{Common evaluation metrics for recommender systems}
\label{tab:recommender-metrics}
\begin{tabularx}{\textwidth}{
    >{\raggedright\arraybackslash}p{2.1cm}
    >{\raggedright\arraybackslash}p{2.3cm}
    >{\raggedright\arraybackslash}X
    >{\raggedright\arraybackslash}p{1.6cm}
}
\toprule
\textbf{Metric} & \textbf{Task} & \textbf{Interpretation} & \textbf{Better value} \\
\midrule
RMSE & Rating prediction & Root mean squared difference between predicted and true ratings & Lower \\
MAE & Rating prediction & Average absolute difference between predicted and true ratings & Lower \\
Precision@K & Top-N ranking & Fraction of recommended items in the top-K list that are relevant & Higher \\
Recall@K & Top-N ranking & Fraction of relevant items retrieved in the top-K list & Higher \\
nDCG@K & Top-N ranking & Ranking quality with higher reward for relevant items at top positions & Higher \\
\bottomrule
\end{tabularx}
\end{table}

Table~\ref{tab:recommender-metrics} summarizes the distinction between rating-prediction evaluation and ranking-based evaluation. Rating-prediction tasks are typically evaluated using error-based measures such as RMSE and MAE, whereas ranking tasks are evaluated using ranking-based measures such as Precision@K, Recall@K, and nDCG@K \cite{herlocker2004evaluating, gunawardana2012evaluating, jarvelin2002cumulated}. The choice of metric must therefore match the recommendation task. To make the evaluation metrics more precise, we introduce their standard mathematical definitions below. RMSE is defined as:

\[
\mathrm{RMSE} = \sqrt{\frac{1}{|T|}\sum_{(u,i)\in T}(\hat{r}_{ui} - r_{ui})^2}
\]

where \(T\) denotes the set of evaluated user-item pairs, \(r_{ui}\) is the true rating, and \(\hat{r}_{ui}\) is the predicted rating. RMSE penalizes larger prediction errors more strongly because the deviations are squared \cite{gunawardana2009survey}. MAE is defined as

\[
\mathrm{MAE} = \frac{1}{|T|}\sum_{(u,i)\in T}\lvert \hat{r}_{ui} - r_{ui} \rvert
\]
and measures the average absolute deviation between predicted and observed ratings \cite{gunawardana2012evaluating}. \\

For Top-N ranking, Precision@K and Recall@K are defined as follows: \\

Precision@K measures the proportion of recommended items in the top-\(K\) list that are relevant, whereas Recall@K measures the proportion of all relevant items that are retrieved within the top-\(K\) recommendations \cite{trattner2023evaluating}:

\[
\mathrm{Precision@K} = \frac{|Rel_u \cap Rec_u@K|}{K}
\]

\[
\mathrm{Recall@K} = \frac{|Rel_u \cap Rec_u@K|}{|Rel_u|}
\]

Here, \(Rel_u\) denotes the set of relevant items for user \(u\), and \(Rec_u@K\) is the set of top-\(K\) recommended items.\\

In contrast to Precision@K and Recall@K, nDCG@K also takes the ranking positions of relevant items into account and can incorporate graded relevance. It is based on Discounted Cumulative Gain (DCG), which rewards relevant items more when they appear near the top of the recommendation list and discounts their contribution at lower ranks. DCG@K is defined as

\[
\mathrm{DCG@K} = \sum_{i=1}^{K} \frac{rel_i}{\log_2(i+1)}
\]

where \(K\) is the cut-off rank and \(rel_i\) denotes the relevance score of the item at position \(i\). The logarithmic discount reduces the contribution of relevant items that appear further down the ranking. This reflects the intuition that highly relevant items should ideally be ranked first and that relevant items appearing lower in the ranking should contribute less to the overall score \cite{gunawardana2012evaluating, jarvelin2002cumulated}.

To make the score comparable across users or recommendation lists, DCG is normalized by the ideal discounted cumulative gain, denoted as \(\mathrm{IDCG@K}\), which corresponds to the best possible ranking up to position \(K\):

\[
\mathrm{IDCG@K} = \sum_{i=1}^{K} \frac{rel_i^*}{\log_2(i+1)}
\]

Here, \(rel_i^*\) denotes the relevance score at position \(i\) in the ideal ranking. The normalized discounted cumulative gain is then defined as

\[
\mathrm{nDCG@K} = \frac{\mathrm{DCG@K}}{\mathrm{IDCG@K}}
\]

Thus, nDCG@K measures how close the actual ranking is to the ideal ranking at cutoff \(K\). Higher values indicate better ranking quality, and with non-negative relevance scores the value typically lies between 0 and 1. Compared to Precision@K and Recall@K, nDCG@K is more informative when not all relevant items are equally important, because it combines both the degree of relevance and the rank position of the recommended items \cite{herlocker2004evaluating, gunawardana2012evaluating, jarvelin2002cumulated}.

%%%%%%%%%%%%%%%%%%%%%%%%%%%%%%%%%%%%%%%%%%%%%%%%%%%%%%%%%%%%
\subsection{Hyperparameter Optimization}

Hyperparameter optimization e selects parameter configurations that improve model performance. Hyperparameters control the behavior of an algorithm, are defined before the training, and not learned from the data. For Example, the number of neighbors in a k-nearest neighbor model or regularization parameters in matrix factorization methods. The choice of hyperparameters can significantly influence model performance \cite{Bergstra2011}.

The hyperparameter optimization problem can be written as the search for a configuration $x$ within a predefined search space $X$ that maximizes or minimizes an objective function: 

\begin{equation}
x^* = \arg\min_{x \in X} f(x)
\end{equation}

Where $f(x)$ is the performance of the model evaluated using a specific metric like nDCG or recall. In practice, the objective function corresponds to the value of the selected evaluation metric computed on a validation dataset. Evaluating a configuration is computationally expensive because it requires training and validating a model on a dataset \cite{Feurer2019AutoML, beel20224}.

The search space is defined by the user and specifies all possible hyperparameter configurations. Each hyperparameter is associated with a range of valid values, which may be continuous, discrete, or categorical. In recommender systems, some hyperparameters depend on the selected algorithm. Such dependencies lead to structured or conditional search spaces, which increases the complexity of the optimization problem \cite{Hutter2011SMAC}.

The optimization budget determines how many configurations can be evaluated during the search process. It can be Specified as a maximum number of evaluations, a time limit, or available computational resources. Since each configuration must be evaluated by training and validating a model, the computational cost of this evaluation limits the number of configurations that can be explored. Efficient optimization methods aim to identify high-performing configurations within a limited budget \cite{Snoek2012BO, wegmeth2025green, arabzadeh2024green}.

Several optimization strategies are used to explore the search space. Grid search evaluates configurations on a predefined grid of parameter values. This approach is simple, but it does not scale well in high-dimensional spaces. Random search instead samples configurations randomly and has been shown to be more effective in high-dimensional settings because it explores a wider range of configurations \cite{Bergstra2011}

More advanced methods use probabilistic models to guide the search process. Bayesian optimization constructs a surrogate model of the objective function and uses this model to select promising configurations for evaluation. This approach balances the exploration of unknown regions of the search space with the exploration of regions that are expected to yield good performance \cite{Shahriari2016BO}. Bayesian optimization methods have been successfully applied in AutoML and automated recommender system pipelines to improve the efficiency and reproducibility of hyperparameter tuning \cite{vente2023introducing, vente2023advancing}.

Hyperparameter optimization plays a central role in modern machine learning pipelines. This combination of complex search spaces, expensive objective functions, and limited optimization budgets motivates the use of automated optimization methods. These methods enable systematic exploration of configurations and reduce the reliance on manual experimentation \cite{vente2025potential, vente2024clicks}.

\subsection{AutoRecSys and the CASH Problem}

Automated recommender systems (AutoRecSys) transfer the core idea of Automated Machine Learning (AutoML) to recommender system pipelines \cite{vente2023introducing, zheng2023automl, vente2023advancing}. Instead of manually selecting a recommendation algorithm and then tuning its hyperparameters, an automated recommender system jointly searches over candidate algorithms and their hyperparameter configurations to identify a strong model for a given dataset and recommendation task \cite{vente2025aps, thornton2013auto, hutter2019automated}. This is particularly relevant in recommender systems because algorithm performance strongly depends on the dataset and the recommendation task \cite{zheng2023automl}. Recent AutoRecSys tools follow this idea. For example, Auto-Surprise automates algorithm selection and configuration for the Surprise library, while LensKit-Auto enables automatic algorithm selection, hyperparameter optimization, and ensembling for LensKit models \cite{anand2020auto, vente2023introducing, vente2024greedy}.

A key limitation of standard hyperparameter optimization is that it typically considers the hyperparameters of a given learning algorithm, rather than selecting the algorithm itself \cite{hutter2019automated}. In practice, however, this assumption is often unrealistic, since different algorithms can perform best on different datasets, and their performance can change substantially depending on how they are tuned \cite{thornton2013auto, hutter2019automated}. For this reason, algorithm selection and hyperparameter optimization should be treated as a single joint optimization problem rather than as two separate steps \cite{thornton2013auto, hutter2019automated}.

This joint optimization problem is commonly referred to as the Combined Algorithm Selection and Hyperparameter optimization (CASH) problem \cite{thornton2013auto, hutter2019automated}. Formally, this problem can be expressed as follows. Let \(A = \{A^{(1)}, \dots, A^{(R)}\}\) denote a set of candidate algorithms, and let \(\Lambda^{(j)}\) denote the hyperparameter space of algorithm \(A^{(j)}\) \cite{hutter2019automated}. The goal of CASH is to identify the algorithm \(A^*\) and configuration \(\lambda^*\) that optimize a validation objective \cite{hutter2019automated, thornton2013auto}:
\[
(A^*, \lambda^*) \in \arg\min_{A^{(j)} \in A,\; \lambda \in \Lambda^{(j)}} \frac{1}{K} \sum_{i=1}^{K} L\!\left(A^{(j)}_{\lambda}, D_{\mathrm{train}}^{(i)}, D_{\mathrm{valid}}^{(i)}\right)
\]
where \(L\) denotes the loss or optimization criterion on the validation data \cite{hutter2019automated}. In recommender systems, this criterion can correspond to minimizing prediction error, such as RMSE, or to optimizing ranking quality, such as nDCG@K, depending on the task \cite{gunawardana2012evaluating, herlocker2004evaluating}.

The CASH problem is challenging because the search space is large, heterogeneous, and often hierarchical \cite{thornton2013auto, hutter2019automated}. Different algorithms expose different hyperparameters, many of which are only meaningful if a particular algorithm has been selected \cite{thornton2013auto}. As a result, the search space contains conditional dependencies, categorical and continuous variables, and potentially many dimensions \cite{thornton2013auto, hutter2019automated}. In recommender systems, this challenge is further complicated by expensive model evaluation, sparse interaction data, and the need to optimize task-specific evaluation metrics.

LensKit-Auto addresses this challenge by extending LensKit with automatic algorithm selection and hyperparameter optimization for recommender system experiments \cite{vente2023introducing}. It supports automated search over candidate recommender algorithms and their configurations for both rating-prediction and Top-N recommendation tasks \cite{vente2023introducing}. In this sense, LensKit-Auto can be understood as a practical AutoRecSys toolkit that addresses a recommender system specific instance of the CASH problem as part of a broader automated recommendation pipeline.

\subsection{Tree Parzen Estimator}
\label{sec:tpe}

Among Bayesian optimization methods for hyperparameter optimization, the Tree Parzen Estimator (TPE) provides an approach based on density estimation \cite{Bergstra2011}. Unlike Gaussian process-based Bayesian Optimization methods, which directly model the objective function, TPE models probability densities over hyperparameter configurations. This formulation enables efficient optimization in search spaces that include categorical variables, conditional parameters, and hierarchical structure \cite{vente2024clicks}. 

Instead of modeling the objective function $f(x)$, TPE reformulates the optimization problem by modeling the conditional probability $p(x \mid y)$, where $x$ represents a hyperparameter configuration and $y$ denotes the observed performance metric. Using Bayes’ theorem, the expected improvement acquisition function can be expressed as follows:

\begin{equation}
p(x \mid y) = \frac{p(y \mid x)p(x)}{p(y)}
\end{equation}

During optimization, the algorithm maintains a history of previously evaluated configurations and their corresponding objective values. TPE partitions the observed configurations into two groups based on a threshold value $y^*$. The threshold value $y^*$ is typically determined using a quantile parameter which specifies the fraction of observations considered promising. Formally, the threshold is defined such that:

\begin{equation}
P(y < y^*) = \gamma
\end{equation}

Where $\gamma$ controls the trade-off between exploration and exploitation. Smaller values of $\gamma$ focus on searching a smaller set of high-performing configurations, while larger values encourage broader exploration of the search space. Configurations that produce good performance ($y < y^*$) form one group, while the remaining configurations form another group. Two density models are then constructed: 

\begin{equation}
l(x) = p(x \mid y < y^*)
\end{equation}

\begin{equation}
g(x) = p(x \mid y \geq y^*)
\end{equation}

The distribution $l(x)$ represents hyperparameter configurations associated with good performance, whereas $g(x)$ represents configurations associated with worse performance. These densities are estimated using non-parametric density estimation techniques such as Parzen window estimators \cite{Parzen1962}.

The acquisition strategy used by TPE selects new hyperparameter configurations that maximize the ratio between the two densities:

\begin{equation}
\frac{l(x)}{g(x)}
\end{equation}

Configurations with large values of this ratio are more likely to belong to the region of the search space associated with better performance. This formulation can be shown to be equivalent to maximizing the expected improvement acquisition function under the TPE model \cite{Bergstra2011}. In practice, TPE generates a set of candidate configurations by sampling from density $l(x)$. The algorithm then evaluates the ratio $\frac{l(x)}{g(x)}$ for these candidates and selects the configuration with the highest value. This sampling-based procedure allows the optimizer to efficiently explore promising regions of the search space without requiring direct optimization of the acquisition function \cite{Bergstra2011}.

One advantage of the TPE approach is that it supports complex search spaces that include conditional and categorical search parameters. Many hyperparameter optimization problems involve hierarchical parameter structures in which the presence of one parameter depends on the value of another parameter. Tree-based representation allows TPE to naturally model such dependencies, which motivates the name of the algorithm. This flexibility makes the method well suited for machine learning pipelines that contain multiple algorithm components with different parameter types. 

Another advantage of TPE lies in its scalability. Because it relies on density estimation rather than global function modeling, the computational cost grows more slowly with the number of observations compared to Gaussian process-based methods \cite{Bergstra2011}. This property makes TPE suitable for optimization problems involving large search spaces or many evaluations.

TPE has been successfully applied to several optimization frameworks. The algorithm forms the basis of the Hyperopt library and has been used in automated machine learning systems to optimize model configurations across different machine learning tasks \cite{Bergstra2011, Feurer2015AutoSklearn}. Due to its ability to handle mixed parameter types and complex search spaces, TPE remains a widely used optimization method for hyperparameter tuning.

\subsection{Meta-Learning in AutoRecSys}
\label{sec:meta}

Automated recommender systems (AutoRecSys) extend the principles of automated machine learning to the domain of recommender systems, with the goal of making the design and optimization of recommendation models accessible to non-experts. Similar to general AutoML frameworks, AutoRecSys systems automate different steps of the recommender system pipeline, such as pre-processing of the input data, choosing a suitable recommendation algorithm, or performing hyperparameter optimization. Such systems typically apply techniques such as Grid Search, Random Search or more advanced optimization strategies like Bayesian Optimization to explore the search space of candidate models to find a well-performing model configuration for the provided input data. However, these methods are computationally expensive because they require training and evaluating many candidate models on the input data \cite{meta_learning, metaLearnerPaper, meta_learning_recsys, wegmeth2024recommender}.

Meta-Learning addresses this problem by using information gained from previous recommender system datasets to improve performance on new datasets. Instead of starting the optimization process completely from scratch for every new problem, a system employing meta-learning uses information about different datasets, algorithms, and their performance on those datasets to guide the exploration of the search space. Instead of training and evaluating many or all available candidate algorithms on a new dataset, a \emph{meta-model} is trained to predict the expected performance of those algorithms based on characteristics of the dataset. The goal is to find the most suitable algorithm -- or a ranked list of algorithms -- for a new dataset without having to exhaustively search the space of candidate algorithms. These promising algorithms can then be prioritized early in the search process, or you can skip the search process altogether. This greatly reduces the computational complexity of the algorithm selection and hyperparameter optimization steps of the AutoRecSys pipeline \cite{meta_learning, metaLearnerPaper, meta_learning_recsys}.

In order to employ Meta-Learning, one must first collect the meta-data which is used to train the meta-model. The first step is acquiring sufficiently many datasets from the domain of recommender systems, such as MovieLens \cite{movieLens} or Amazon \cite{amazon} datasets. Then you extract from the collected datasets those characteristics, \emph{meta-features}, that will later be used to train the meta-model on. Which characteristics of the datasets are chosen as the meta-features depends on the application and aims of the specific AutoRecSys framework. Such meta-features could include the number of users and items, the number of interactions, the sparsity of the interaction matrix, or the mean item rating \cite{meta_learning, metaLearnerPaper, meta_learning_recsys}.

Once you collected enough datasets and extracted their meta-features, the next step is to gather how well each available algorithm performs on each collected dataset. If this information is not already available, one would need to train each algorithm on each dataset and evaluate the performances using appropriate metrics like NDCG or Recall. The meta-features of each dataset and the performances of each algorithm on that dataset is what the meta-model will be trained on. The meta-model is trained to predict either the performance score of each algorithm or a ranking of the top-N algorithms on a new unknown dataset. The meta-model acts as a black box, because it has no information on how the algorithms function or their concrete implementations. It also knows nothing about what information each dataset actually represents. The only information the meta-model has to make inferences on algorithm performance on new datasets is the meta-features of each dataset and how well each algorithm performs on those datasets \cite{meta_learning_blackbox}. It is assumed that if two datasets are very similar in terms of their meta-features, the performance of each algorithm on those datasets will also be very similar \cite{meta_learning, metaLearnerPaper, meta_learning_recsys}.

The trained meta-model can now be integrated into the AutoRecSys framework. Whenever the user provides a new dataset to the framework, its meta-features are extracted and fed to the meta-model, which will then infer which algorithms are expected to perform well on that dataset. This result can then be used either to directly select the most promising algorithm and continue with hyperparameter optimization, or to provide the basis for a smaller search of the space of candidate models. Since algorithms that are likely to perform poorly on a given dataset are disregarded early on in the AutoRecSys pipeline, the search space of candidate models and therefore the computational complexity of the algorithm selection step is greatly reduced \cite{meta_learning, metaLearnerPaper, meta_learning_recsys}.

In Section \ref{sec:meta-learner} we describe the adaptation of an existing meta-learning framework and the acquisition of a meta-dataset suitable for LensKit-Auto.

\section{LensKit-Auto Framework}
LensKit-Auto\footnote{https://github.com/ISG-Siegen/lenskit-auto} is an experimental Automated Recommender System (AutoRecSys) toolkit based on the LensKit \cite{ekstrand2020lenskit} framework. LensKit-Auto extends the LensKit framework with functionality for automated recommender system experimentation. In particular, it automates algorithm selection, hyperparameter optimization, and post-hoc model ensembling for both rating-prediction and top-N recommendation scenarios \cite{vente2023introducing}. In this way, LensKit-Auto reduces the manual effort required to configure, compare, and evaluate recommender models.

\begin{figure}[t]
    \centering
    \includegraphics[width=0.95\textwidth]{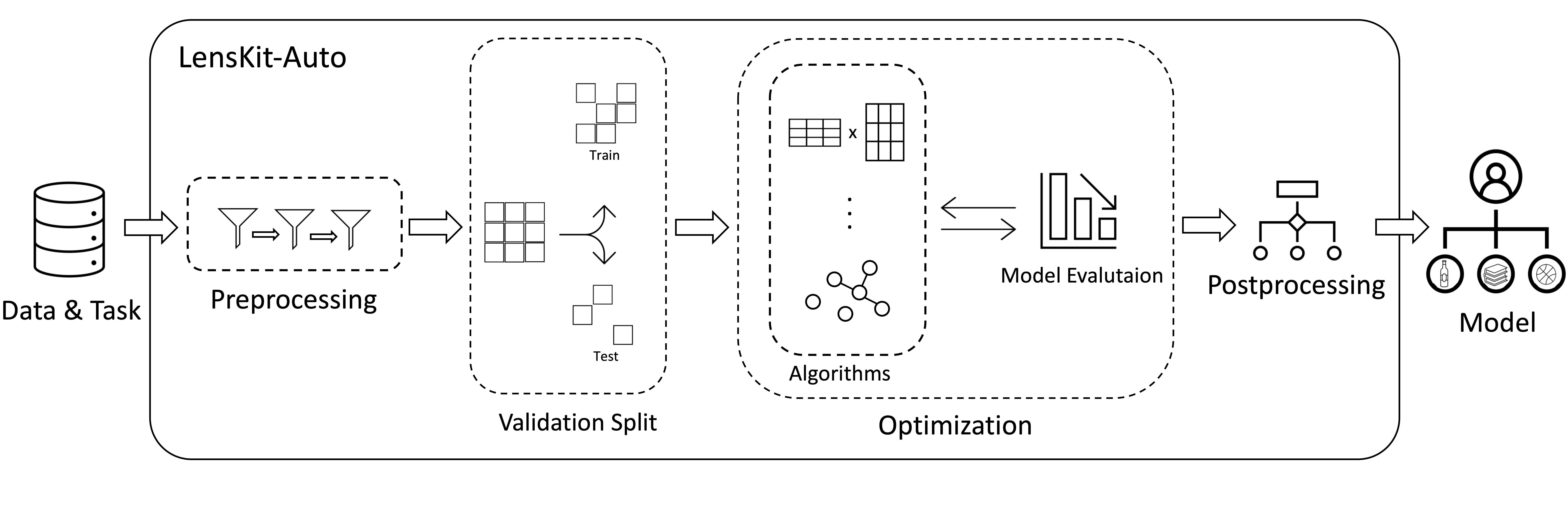}
    \caption{High-level workflow of LensKit-Auto.}
    \label{fig:lenskit-auto-pipeline}
\end{figure}

At a high level, LensKit-Auto follows an automated workflow that takes a dataset together with a recommendation task and returns a recommender model selected through the optimization process. To achieve this, the framework structures model selection and evaluation into several stages, including pre-processing, validation splitting, optimization, model evaluation, and postprocessing. Candidate models are generated and evaluated during this workflow, and the final output is the best-performing recommender model for the chosen task (rating-prediction or Top-N recommendation). Figure~\ref{fig:lenskit-auto-pipeline} from \cite{vente2023introducing} illustrates the high-level workflow of LensKit-Auto, from the input dataset and task definition to the final optimized model.

The pre-processing stage prepares the dataset later used in the experimentation. In this stage, duplicate rows and rows containing \texttt{NaN} values (missing values) are removed. In addition, users with very few or very many interactions can be filtered out by using the parameters \texttt{min\_interactions\_per\_user} and \texttt{max\_interactions\_per\_user}.

After pre-processing, LensKit-Auto creates validation splits using \texttt{validation\_split(...)}.  Depending on the selected setting, this can produce either user-based or row-based holdout splits. These splits are then used by the evaluator classes in \texttt{*\_evaler.py}, which train candidate models on the training portion of each fold and evaluate them on the corresponding validation portion. The resulting validation scores are then passed to the optimization component, where different candidate configurations are compared.

The optimization components are the core of LensKit-Auto. In this phase, the framework searches over recommender algorithms and their hyperparameter configurations. For each sampled configuration, a corresponding LensKit model is build and evaluated according to the selected recommendation task and metric. The optimization process therefore links together several parts of the system: the configuration space, the recommender algorithms, the evaluators, and the metric-driven comparison of results. In the original framework, random search \cite{bergstra2012random} and Bayesian optimization \cite{snoek2012practical} already formed the main search strategies, while the project described in this report extends this functionality further in Section \ref{subsection:TPE}.

After optimization, LensKit-Auto performs a postprocessing step in order to produce the final output model. In the prediction workflow, this may involve either selecting the best single configuration or building an ensemble from the top-performing runs, depending on the chosen \texttt{ensemble\_size}. In the recommendation workflow, postprocessing mainly consists of constructing the final model from the best configuration found during optimization. In the extended version developed in this project, the resulting model and its configuration can additionally be saved, and the hyperparameter optimization runs can be visualized, as described in Chapter~\ref{chap:improvements}.

This Chapter has outlined the original structure and workflow of LensKit-Auto as the foundation of the project. Based on this, Chapter \ref{chap:improvements} describes the concrete extensions and modifications that were implemented, including compatibility fixes, the addition of the Tree Parzen Estimator optimization strategy, DeepCAVE integration, testing improvements, and documentation updates.

\section{Improvements to LensKit-Auto}
\label{chap:improvements}
The goal of this project is to update LensKit-Auto, since LensKit got a major update, and to add additional functionality to LensKit-Auto. In this section, we will take a look at the improvements that were made to LensKit-Auto and how these are implemented.

\subsection{LensKit Framework Update}
\label{sec:lk_update}
LensKit-Auto builds upon the LensKit framework, which received a big update in March, 2025. This led to incompatibilities with the new LensKit versions. The way LensKit interacts with data has fundamentally changed, as has the framework's structure. Like the release notes of LensKit\footnote{https://lenskit.org/stable/releases/2025.html\#id21} already state, there are too many changes to list them all here, so we will just briefly look at the biggest ones in this section.

Some of the changes introduced in the LensKit 2025.x versions are the classes handling data. When LensKit was mainly using Pandas data frames in the past, it now includes its own \texttt{Dataset} class for handling large amounts of data. Recommendations and predictions are now returned as \texttt{ItemLists} and \texttt{ItemListCollections}. Another addition is the \texttt{Pipeline} class, which wraps components. A component's task is to take an input and produce an output, which is exactly the task of a prediction- or recommender-model.

Especially the changes in the data structures affected the code. Many functions needed to be adapted to the new data-sets and item-lists.

Other noteworthy changes that do not influence LensKit-Auto directly, but ease its use, are, for example, the reworked \texttt{load\_movielens} function or the \texttt{RunAnalysis} class. The new \texttt{load\_movielens} function can be used to load any MovieLens dataset into a \texttt{Dataset} from a zip-file. With a run analysis object we can measure the accuracy of predictions from our trained model. These functions substantially ease the process of integrating LensKit-Auto into another project.

In the next sections, we will look at the changes we implemented in LensKit-Auto and how they were achieved.
%===================

\subsection{Implementation of Tree Parzen Estimator}
\label{subsection:TPE}
To extend LensKit-Auto with an additional hyperparameter optimization strategy, the Tree Parzen Estimator (\textbf{TPE}) algorithm was implemented as a new optimization module. While the theoretical foundations of TPE were introduced in Section \ref{sec:tpe}, this section describes the technical implementation and integration of the algorithm into the LensKit-Auto framework. 

\subsubsection{Use of the Hyperopt Library}

The implementation closely follows the approach provided by the Hyperopt framework, which includes the Tree Parzen Estimator Algorithm \cite{Bergstra2011}. Instead of implementing the optimization entirely from scratch, the Hyperopt library was used to handle the underlying TPE optimization logic. 

Hyperopt provides the \texttt{fmin} optimization routine, which iteratively evaluates candidate configurations using a specified optimization algorithm. In the implementation, the \texttt{tpe.suggest} function is used as the search algorithm within \texttt{fmin}. This function generates new hyperparameter configurations based on the density estimation procedure described in Section \ref{sec:tpe}. 

The optimization process is executed as follows:

\begin{enumerate}
    \item A hyperparameter search space is defined using the Hyperopt parameter definition functions. 
    \item An objective function is defined that evaluates a given hyperparameter configuration. 
    \item The \texttt{fmin} function repeatedly samples configurations using the TPE algorithm. 
    \item Each sampled configuration is evaluated using the LensKit evaluation pipeline. 
    \item The configuration that achieves the best performance according to the chosen optimization metric is returned. 
\end{enumerate}

The use of Hyperopt simplifies the implementation and ensures that the TPE algorithm follows a well-tested, widely adopted approach.

\subsubsection{Integration with LensKit-Auto}

The TPE optimizer was integrated as a new optimization strategy within the \texttt{optimization\_strategies} module of LensKit-Auto. The implementation is contained in the file \texttt{tree\_parzen\_estimator.py}, which defines the function that serves as the entry point for performing hyperparameter optimization using the TPE algorithm. 

The TPE optimizer was integrated into the existing optimization pipeline by extending the selection mechanism that’s used in the \texttt{lkauto.py} module. LensKit-Auto already supports random search and Bayesion optimization strategies. To incorporate TPE, a new ''tpe'' option was added to the \texttt{optimization\_strategie} parameter. When this option is selected, the optimization process invokes the \texttt{tree\_parzen} function, which performs the hyperparameter search using the TPE algorithm. 

This allows TPE to be used interchangeably with existing optimization strategies without requiring changes to LensKit-Auto's external API. Users can therefore select the optimization strategy by specifying the desired algorithm in the configuration parameters.

\subsubsection{Objective Function and Evaluation Pipeline}

The objective function defines how a candidate hyperparameter configuration is evaluated. The objective function receives a sampled configuration from Hyperopt and passes it to the LensKit evaluation pipeline. 

Evaluation is performed using specialized evaluator classes based on the dataset's feedback type. For datasets having explicit ratings, the \texttt{ExplicitEvaler} class is used. For datasets containing implicit feedback, the \texttt{ImplicitEvaler} class is used. These evaluators train the corresponding recommender model and compute the specified evaluation metrics. 

The objective function returns a dictionary which contains the evaluation result in the format required by Hyperopt. The key \texttt{loss} stores the performance value of the evaluated configuration, and the \texttt{status} field indicates successful evaluation. Hyperopt uses this information to update its internal density models and guide the search process toward promising regions of the configuration space. 

\subsubsection{Hyperparameter Search Space Representation}

To enable compatibility with Hyperopt, the hyperparameter search space representation was extended. LensKit-Auto originally defined configuration spaces using the ConfigSpace Library. However, Hyperopt requires search spaces to be expressed using its own parameter definition functions. 

To support multiple optimization strategies, a flag was introduced in the configuration space generation functions. When the \texttt{hyperopt} flag is enabled, the configuration space is defined using Hyperopt parameter distributions such as \texttt{hp.uniform}. Otherwise, the  ConfigSpace representation is used.  

This design enables LensKit-Auto to maintain compatibility with multiple optimization strategies while minimizing changes to the existing configuration space definitions. 

\subsubsection{Optimization Workflow}

The overall optimization workflow of the TPE implementation follows the standard procedure used by Hyperopt. First, the search space and objective function are defined. A \texttt{Trials} object is then created to store the history of evaluated configurations. The \texttt{fmin} function repeatedly samples configurations using the TPE algorithm until the maximum number of evaluations is reached. 

During optimization, the evaluator computes the chosen performance metric for each configuration. The optimizer tracks the best configuration observed during the search process. After optimization completes, the optimized configuration and evaluation results are returned to the main LensKit-Auto pipeline. 

For explicit feedback tasks, the evaluation results are additionally stored as a DataFrame containing the top-performing configurations. These results can be used for ensemble construction or further analysis of the optimization process. 

\subsubsection{Result Handling and Model Construction}

Once the optimization procedure finishes, the best hyperparameter configuration is used to build the final recommender model. The LensKit-Auto pipeline converts the configuration into the corresponding LensKit model using the \texttt{get\_model\_from\_cs} utility function. 

If ensemble learning is enabled, the top-performing configurations can be combined to form an ensemble model using the \texttt{build\_ensemble} function. Otherwise, the single best configuration is used to create the final model. 

The resulting model and configuration are optionally saved to disk using the Filer utility. This allows trained models and optimization results to be reused in later experiments.

%============================

\subsection{Model Persistence}
In the earlier version of LensKit-Auto, it was possible to find the best recommender or prediction model for the users use case, along with the best fitting values for its hyperparameters. The drawback is that the output of LensKit-Auto contained only this information, but not the actual trained model. So the result of a LensKit-Auto run could look like this:

{\texttt{INFO \{'algo': 'ItemItem',}\\
\null\qquad\qquad\texttt{'ItemItem:min\_nbrs':  10,}\\
\null\qquad\qquad\texttt{'ItemItem:min\_sim':  0.00169,}\\
\null\qquad\qquad\texttt{'ItemItem:nnbrs':  9043\}}}

\noindent While this information is very useful, it would be even better if a LensKit-Auto run returned the trained model that works best for the users data set. So in the new version of LensKit-Auto, the trained model is returned at the end of the run, as well as saved to the hard drive in a pkl-file. Saving to the hard drive is only enabled if the parameter \texttt{save} in the function-call is set to \texttt{True}.

The implementation is straightforward. During the optimization, different models with different hyperparameters are evaluated. The current best performing model is stored in a variable, which is updated if a new model-hyperparameter-configuration is found that has a lower cost. At the end of a LensKit-Auto run, this stored model is returned to the user and stored in a pkl-file using Python's pickle library. Together with the model, the incumbent gets returned and saved too. The previous print-output is kept because it contains the crucial information on what kind of model is actually returned.

These additions to LensKit-Auto improve its usability by allowing the user to reuse the model that was found by LensKit-Auto. Instead of having to train the model ourselves, we now have a model ready to use.

%=====================

\subsection{Limiting the run time}
\begin{lstlisting}[language=Python, frame=single, float=t,
caption={A simplified view of the old random-search implementation.}, label={lst:OldRandomSearch}]
start_time = time.time()
while time.time() - start_time < time_limit:
    # pick configuration
    error, model = evaler.evaluate(configuration)
    if error < best_error:
        best_error = error
        best_configuration = configuration
        best_model = model
return best_configuration, best_model
\end{lstlisting}

\noindent Another improvement to LensKit-Auto is that the time-limit, which can be set when starting the program, is now more strictly enforced by the random-search. An abstraction of the random search implementation in the old LensKit-Auto version is shown in listing \ref{lst:OldRandomSearch}. As we can see, the elapsed time is only checked after every iteration of the evaluator evaluating the randomly picked configuration. This can lead to longer-than-expected run times if the evaluation of a configuration takes especially long. To fix this, we introduce Python processes.

To implement the enforcement of the time-limit, we first need to calculate the timestamp at which the random search should stop. The sum of the current timestamp and the \texttt{time\_limit\_in\_sec} parameter that is passed to the \texttt{random\_search} is the stopping timestamp. The first step is to outsource the code block that follows after a configuration is picked (the code after \texttt{\# pick configuration} in listing \ref{lst:OldRandomSearch}). To run this code block as a process, we outsource it into a function we call \texttt{evaler\_worker\_function}. In this worker-function the evaluator is called to retrieve the error and model of the configuration. If the current error is better than the previous best error, the model and error get stored as the new benchmark to beat. To access the results of the worker-function, we also need to create a queue from \texttt{multiprocessing.Queue} that is passed to the worker-function. At the end, the results, such as the best model and best error score, are put into the queue. 

The random-search function now iterates over a set of configurations and randomly selects one. It creates a process that runs the worker-function, evaluates these configurations and their models, and puts the results into the queue. In the random-search function we can try to access the queues contents with \texttt{queue.get()} and afterwards join the process. The \texttt{join()} method takes in the parameter \texttt{timeout}, which is used to set a time after which the process is forced to join, if it is done executing or not. The timeout is set to the time that is left for the optimization. This leads to the possibility that the queue can be empty. To counteract this, this whole operation is wrapped up in a try-except-block, with the exception being \texttt{multiprocessing.queues.Empty}. If the queue is empty and the process is still alive, it is terminated.

If we set the time-limit to e.g. 120 seconds, the random-searches execution time is now much closer to the time-limit than before. Though usually it is a little over the limit, since the function includes some overhead like terminating and safely joining the process.

This optimization only works for the random-search. The bayesian optimization uses SMAC, a third party library, whose contents we don't want to interfere with. SMAC has its own option to specify a time-limit, which in our experience does not overshoot the time-limit too much.

%====================

\subsection{Visualization with DeepCAVE}
\begin{figure}
    \centering
    \subfloat[Configuration cost - performance \label{fig:ConfigCostPerf}]{\includegraphics[width=0.45\linewidth]{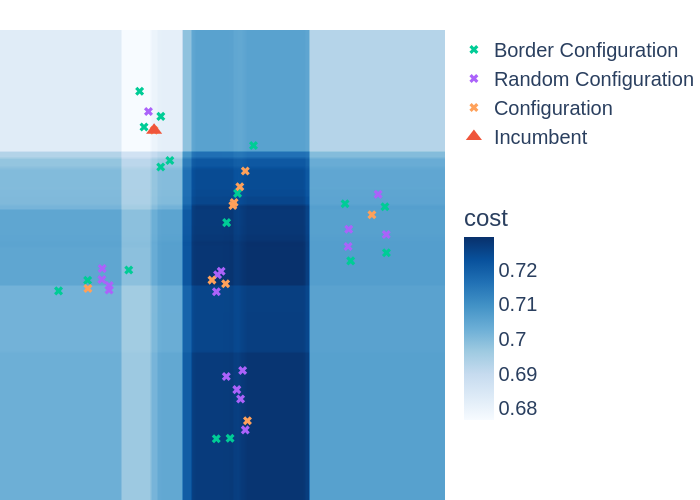}}
    \subfloat[Configuration cost - coverage \label{fig:ConfigCostCov}]{\includegraphics[width=0.45\linewidth]{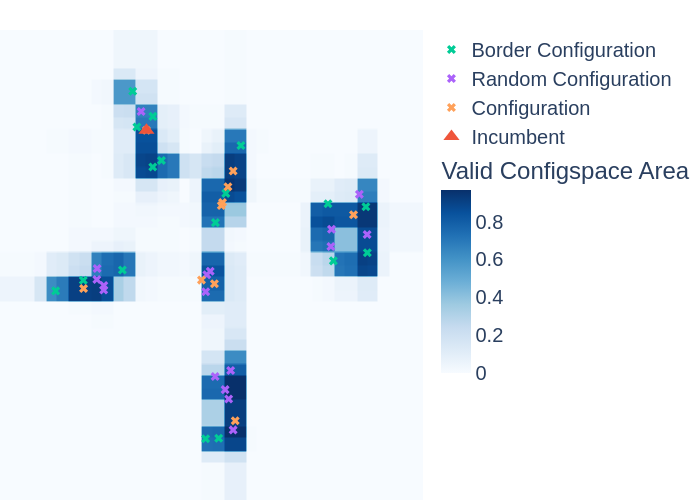}}
    \\ %linebreak
    \subfloat[Pareto front \label{fig:ParetoFront}]{\includegraphics[width=0.45\linewidth]{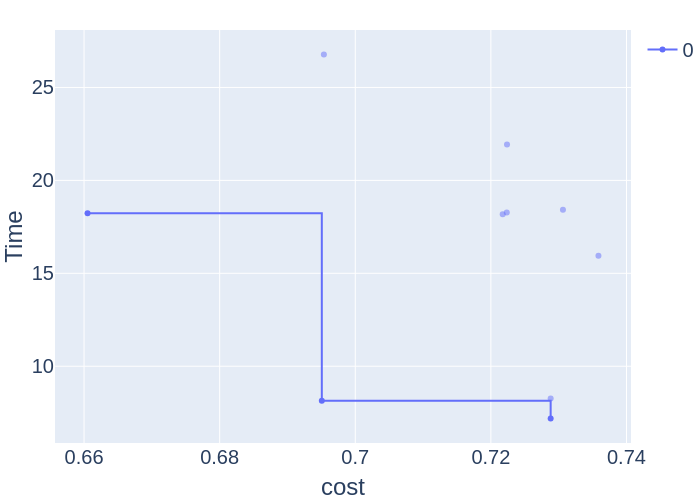}}
    \subfloat[Cost over time \label{fig:CostOverTime}]{\includegraphics[width=0.45\linewidth]{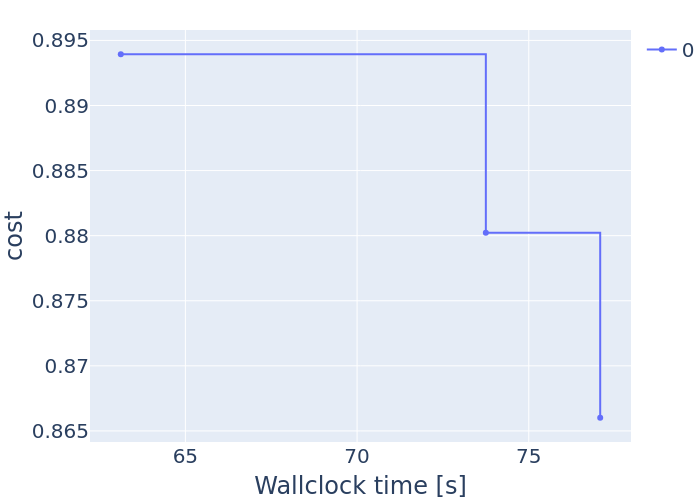}}
    \caption{Some examples of the DeepCAVE output}
    \label{fig:deepcave}
\end{figure}
To visualize the optimization process of LensKit-Auto, we added the DeepCAVE library. DeepCAVE is a framework that generates tables and graphics for AutoML systems. It focuses on the hyperparameter optimization aspect of AutoML systems and can help to easier understand the optimization process. For example, the configuration-footprint figure can help to understand how much of the configuration space was explored \cite{deepcave}. This fits our use case nicely.
DeepCAVE provides a lot of possible figures. In this section we will look at the ones we chose to implement into LensKit-Auto, and why.

\paragraph{Cost over time}
The Cost over Time figure is pretty self-explanatory. It shows the course of the cost on the y-axis and the elapsed time on the x-axis. Note that not all configurations get displayed. We included this figure because it highlights what the highest and lowest found costs are, but normally the true information is best found when we load the DeepCAVE UI in a browser to inspect the graph. Here we can hover over the dots on the graph which reveals information about the configuration, like the algorithms name and the values of its hyperparameters. 

But, while using the tool ourselves, we found a bug in DeepCAVE. While inspecting the graph, we noticed that some configurations are labeled wrong. After reading the runhistory file created by SMAC, it is clear that the information revealed when hovering over a dot is correct (i.e. the cost, time and hyperparameter values are correct and belong to the run with that \textit{Configuration ID}), but they are plotted at the wrong position. The graphic we are considering can be seen in \ref{fig:CostOverTime}. In our case, the dot at the top with the highest cost (around 0.894) is labeled with the configuration with best lowest cost score (around 0.865). That matches with the position of the dot at the "bottom right" in the graph. The label information for that dot at the bottom also matches with the values of the dot at the top. So in short, the trend-line of the cost is still correct, but the labels for the best and worst configurations are switched up. So while the graphic can still be useful, the further information for the configurations should be taken with care and be cross checked. 

\paragraph{Pareto Front}
A pareto front figure highlights configurations that show a balance between two optimization factors \cite{deepcaveDocs}. With the elapsed time on the y-axis and the cost on the x-axis, we can choose a configuration that fits our needs. The configuration with the lowest cost might take the longest to train and the configuration that is the fastest to train might have the highest costs. In some scenarios a configuration between these two could be the one that fits our needs. A configuration that does not lay on the pareto front is not the best option regarding the tradeoff between cost and time. The configuration in Figure \ref{fig:ParetoFront} with a cost around 0.695 and time 27 has a lower cost than other configurations, but the configuration beneath it on the pareto front has the same cost with a lower training time.

This figure shows that there are other configurations that, for example, might have the same cost as another one, but are worse in another aspect (the time). But again, more information is revealed in the UI version of DeepCAVE, where we can see what the configurations values are.

\paragraph{Configuration Footprint}
The configuration footprint figures visualize how well the LensKit-Auto run covered the possible configuration space. The figure plots different kind of configurations in one figure. The \textbf{incumbent}, the best configuration, is marked by a red triangle. Configurations that were \textbf{evaluated} by LensKit-Auto are depicted as an orange \textit{x}, while configurations that were \textbf{randomly sampled} are a purple \textit{x}. \textbf{Border configurations} are configurations that minimize and maximize the possible values in the configuration. These configurations are displayed as a green \textit{x} \cite{deepcaveDocs}. DeepCAVE provides us with two types of configuration footprint figures:
\begin{enumerate}
    \item \textbf{Coverage plot:} The coverage plot works well for highlighting the areas in which LensKit-Auto explored the configuration space \cite{deepcaveDocs}. As we can see in Figure \ref{fig:ConfigCostCov}, we will end up with clusters of configurations. The border configurations are useful here, in the sense that they help us to imagine the boundaries of the configuration space.
    \item \textbf{Performance plot:} The performance plot maps the configurations on top of an objective score map \cite{deepcaveDocs}, in our case the cost. Note that the color for the cost may not be correct for areas where no configuration was evaluated, since these areas are just approximated. The incumbent should lay in the lightest spot, as it does in Figure \ref{fig:ConfigCostPerf}.
\end{enumerate}

\subsubsection{Excluded figure types}
As mentioned before, for some figure types it is a good idea to also load the DeepCAVE user-interface in a browser to inspect the figures, since the UI can display additional information. We deemed some figure types as only being interesting in the browser, because they need the interaction-aspect of the UI. Some figure types also did not produce any meaningful output for us, like the Importances figure. The Importances figure is one of a few figure types, where we can select multiple hyperparameters, that we want displayed in the figure, from the whole range of hyperparameters present in a LensKit-Auto run. This leads to too convoluted figures. We could theoretically generate multiple Importances figures, where each figure represents one algorithm and displays its hyperparameters, but since it would then be possible that a big multitude of images is generated, we decided against it, to keep a more clear overview.

There are also some other figure types that are implemented into the LensKit-Auto output that are not mentioned here. The reason for that is that it is generally possible to implement them, but they often produce an output that contains no real value. E.g. the ablation paths plugin, that can be used to study the effect a hyperparameter change from the default configuration has on the objective \cite{deepcaveDocs}. For some runs we were able to generate outputs with a graph that showed the cost sinking the more hyperparameters are fitted to the incumbent value, but most often than not, the figure shows a blank graph. We kept the implementation anyway, since sometimes the output is viable.

\subsubsection{Conflicting Version Dependencies}
LensKit-Auto uses Python version 3.12, but the DeepCAVE package lists a Python version smaller than 3.11 as a requirement. This is the reason why the DeepCAVE package has to be installed manually after LensKit-Auto, because this way we can add a flag to the pip-install-command that tells pip to ignore the python version requirements.

DeepCAVE itself also installs some requirements, one of them being an older matplotlib version. This also needs to be manually upgraded to the 3.10.3 version afterwards. DeepCAVE also installs the packages Plotly and Kaleido, though the versions DeepCAVE installs are incompatible with each other, so that Kaleido needs to be downgraded to an earlier version, as is shown in the installation instructions.

We added the DeepCAVE functionality to LensKit-Auto anyways, because we believe that it is a nice addition to have, and also because in our experience, DeepCAVE works with newer Python version, like 3.12. We did not run into any problems using DeepDAVE and since it is part of the postprocessing, if there is a problem due to the conflicting versions, it should not influence the optimization process that is already finished once DeepCAVE is started.

\subsection{Improvement of the Testing Infrastructure}

A central improvement of this project concerned the testing infrastructure of LensKit-Auto. The original test suite already contained some tests, but several parts of the codebase were still insufficiently covered. Therefore, the existing tests were reviewed and updated, and many new tests were added. This increased the number of tests from \textbf{26} to \textbf{146} and raised the coverage to \textbf{98\%}.

The newly added tests also included dedicated tests for \texttt{lkauto.py}, which contains the central entry-point functions of the framework. These tests cover the functions \texttt{get\_best\_prediction\_model()} and \texttt{get\_best\_recommender\_model()}, which coordinate pre-processing, optimization, ensemble construction, and model creation. Since these functions connect core components of LensKit-Auto, testing them is especially relevant for verifying the overall behavior.

The new tests verify, among other things, whether the correct optimization strategy is called depending on the selected setting, whether invalid optimization strategies raise the expected \texttt{ValueError}, and whether optional parameters are handled correctly. They also check that ensemble construction is only triggered when \texttt{ensemble\_size > 1}. In addition, the saving behavior is tested by verifying that the \texttt{Filer} utility stores both the trained model and the incumbent configuration when \texttt{save=True} is used. For \texttt{get\_best\_recommender\_model()}, the tests also verify that the optimization is executed with \texttt{predict\_mode=False} and that, if the optimization does not directly return a trained model, a fallback model is correctly constructed from the best configuration using \texttt{get\_model\_from\_cs()}. Altogether, these tests improve coverage of the control flow of LensKit-Auto, and help ensure that the main functions behave correctly under different settings.

Another important focus was the testing of optimization-related functionality. Tests were added and extended for the optimization modules, in particular for Bayesian optimization and random search. In \texttt{random\_search.py}, for instance, the tests cover updated logic, invalid feedback types, evaluator creation, worker execution, queue handling, and time-limit-related behavior.

The testing process not only improved the test suite itself, but also helped identify weaknesses in the implementation. During the work on the tests, several issues in the codebase became visible and were corrected. One example was a bug in the handling of validation splits. The output of \texttt{validation\_split(...)} in \texttt{lkauto/explicit/explicit\_evaler.py} was reused, even though it returned an iterator that could only be iterated over once. In addition, the evaluation code incorrectly accessed the training and test data inside the fold loop. Instead of using the current fold, it accessed \texttt{self.train\_test\_splits.train} and \texttt{self.train\_test\_splits.test}. This was corrected by converting the validation splits into a list and by using \texttt{fold.train} and \texttt{fold.test} inside the loop. The essential change is illustrated in Listing~\ref{lst:validation-split-fix}. The previous version accessed the split data through \texttt{self.train\_test\_splits}, whereas the corrected version first stores the validation splits as a reusable list and then accesses the training and test partitions through the current \texttt{fold}.

\begin{lstlisting}[language=Python, frame=single, float=t,
caption={Simplified correction of validation split handling}, label={lst:validation-split-fix}]
# before
for fold in self.train_test_splits:
    train = self.train_test_splits.train
    test = self.train_test_splits.test

# after
self.train_test_splits = list(validation_split(...))
for fold in self.train_test_splits:
    train = fold.train
    test = fold.test
\end{lstlisting}

This correction ensured that the validation splits could be reused safely and that each iteration operated on the correct fold.

Another implementation issue was identified in \texttt{lkauto/utils/get\_model\_from\_cs.py}, where sampled hyperparameter configurations are translated into recommender models. In this part of the framework, sampled configurations were not always applied correctly to the corresponding model constructors. In addition, configuration keys were not normalized consistently. As a result, different sampled configurations could fail to produce the intended differences in model construction. This issue was corrected by revising the handling of configuration keys and by ensuring that algorithm-specific parameters were passed properly to the corresponding model constructors. The implementation was later further adapted to the updated LensKit 2025 API by constructing explicit configuration objects for the supported algorithms.

Overall, the revision of the testing infrastructure improved not only the coverage percentage of LensKit-Auto, but also the robustness of the framework itself. The added and updated tests cover central workflow logic, optimization components, and important utility functions, while also helping to uncover implementation issues that were corrected during development. This creates a more reliable basis for future extensions of LensKit-Auto.

%%%%%%%%%%%%%%%%%%%%%%%%%%%%%%%%%%%%%%%%%%%%%%%%%%%%%
\subsection{Improvement of the Documentation}

Another important part of this project concerned the revision and modernization of the documentation of LensKit-Auto. A major part of this work consisted of updating the examples and usage instructions to the LensKit 2025.2.0 API. The original documentation still contained examples based on an older LensKit version whose API had changed significantly. As a result, the documented examples no longer matched the current implementation and could not be executed as written.

The \texttt{Getting\_Started} guide was substantially updated to reflect the newer LensKit API and the current usage of LensKit-Auto. In particular, the examples for recommendation and prediction workflows were revised, including updated data-splitting examples for user-based and record-based evaluation. This made the introductory material more consistent with the current implementation of the framework.

In addition, the structure of the Sphinx documentation was improved and the navigation was extended with a separate \texttt{Subpackages} section that provides direct access to major package components such as algorithms, ensemble methods, preprocessing, optimization strategies, and utility modules. Empty and unnecessary sections were removed in order to make the generated API documentation cleaner and easier to navigate.

\begin{figure}[p]
    \centering

    \begin{subfigure}[t]{1\textwidth}
        \centering
        \includegraphics[width=\linewidth]{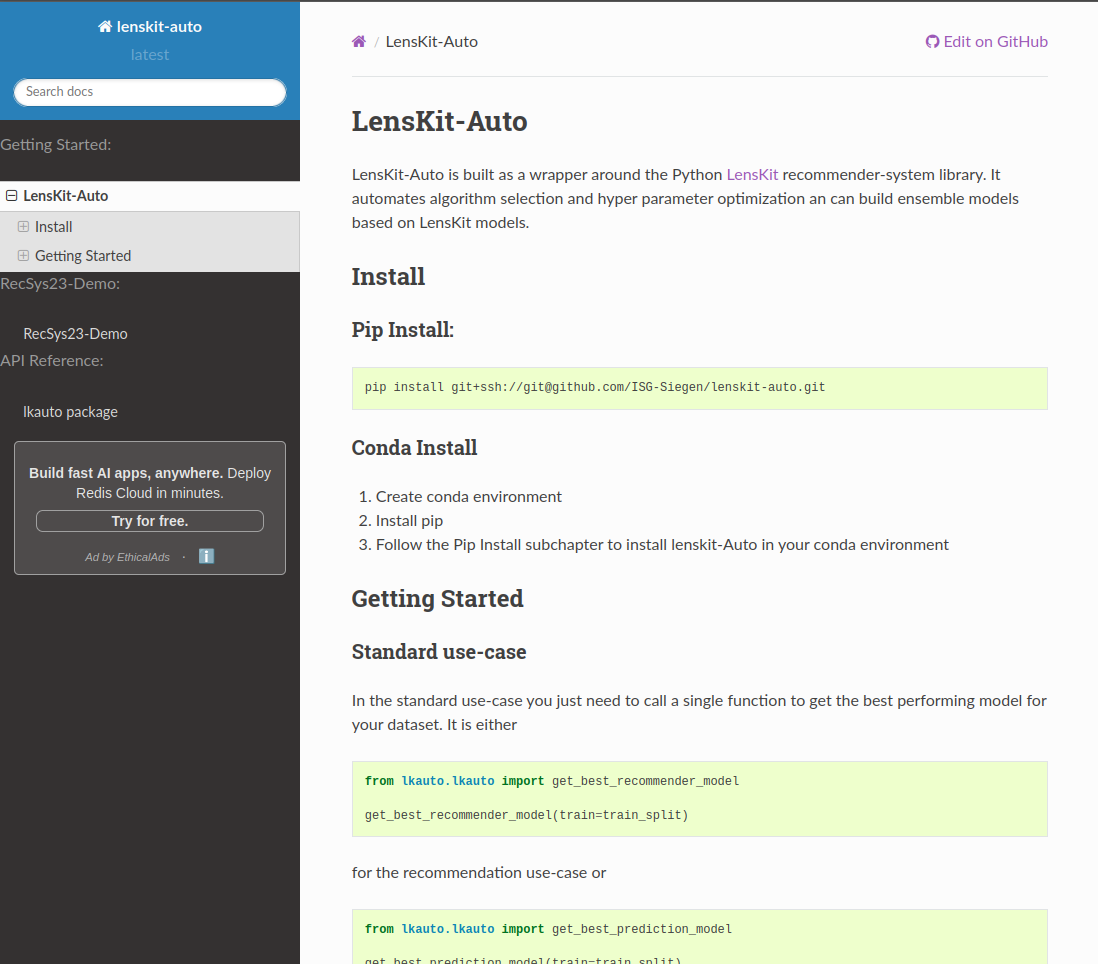}
        \caption{Previous documentation version}
        \label{fig:lkauto-docs-before}
    \end{subfigure}

    \vspace{0.8cm}

    \begin{subfigure}[t]{1\textwidth}
        \centering
        \includegraphics[width=\linewidth]{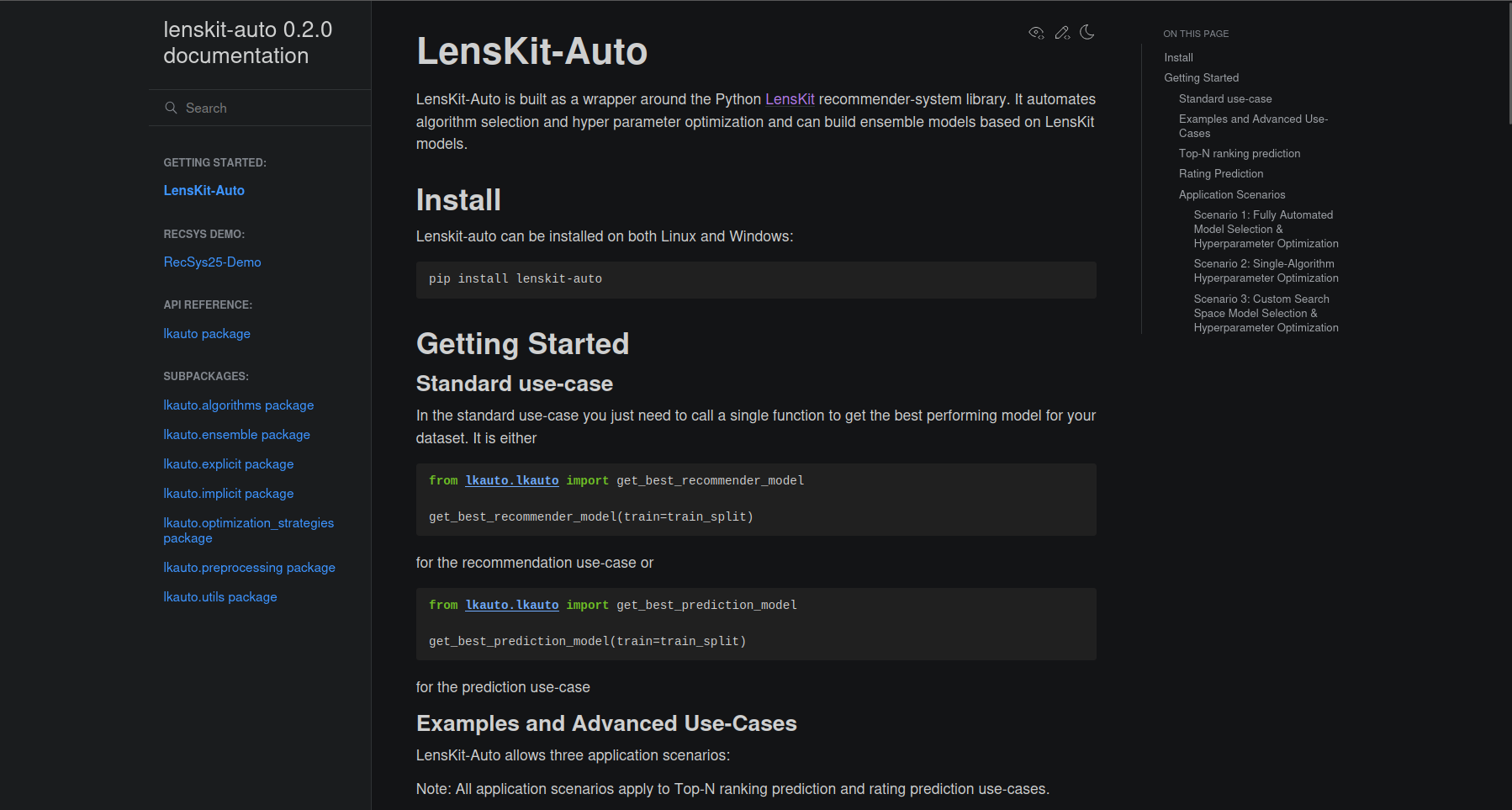}
        \caption{Updated documentation version}
        \label{fig:lkauto-docs-after}
    \end{subfigure}

    \caption{Comparison of the LensKit-Auto documentation before and after the documentation revision.}
    \label{fig:lkauto-docs-before-after}
\end{figure}

The visual effect of these revisions is illustrated in Figure~\ref{fig:lkauto-docs-before-after}, which compares the previous documentation page with the updated version.

Compared with the previous version, the new documentation provides clearer navigation, revised installation instructions, and a more modern presentation. 

The demonstration material was also updated. The older demo notebook was replaced by a new Jupyter notebook that reflects the current LensKit-Auto workflow with the new updated LensKit API. This gives users an executable example that is aligned with the current state of the framework.

Furthermore, the documentation build configuration was updated. Previously, the version number displayed in the documentation had to be updated manually whenever a new release of LensKit-Auto was created. This manual process could lead to mismatches between the package version and the documented version. To address this issue, dynamic versioning was introduced. The documentation now automatically reads the version number from the package variable \texttt{\_\_version\_\_} defined in \texttt{lkauto/\_\_init\_\_.py}. This ensures that changes to the package version are reflected automatically in the documentation without requiring manual updates to the configuration. The documentation theme was also updated to Furo. This updated theme supports both light and dark modes. As shown in the updated documentation view, users can switch between these modes directly via the toggle in the top-right corner of the page.

Finally, the repository-level documentation was improved through updates to the \texttt{README}. Installation instructions were simplified and updated for the current environment requirements, including the \texttt{pip}-based installation, while outdated setup instructions such as the previous SSH-based git installation were removed. The new installation instructions are straightforward and support both Linux and Windows environments:

\begin{verbatim}
pip install lenskit-auto
\end{verbatim}

%%%%%%%%%%%%%%%%%%%%%%%%%%%%%%%%%%%%%%%%%%%%%%%%%%%%%

\subsection{Acquisition of a suitable meta-dataset}
\label{sec:meta-learner}

In Section \ref{sec:meta} we describe how meta-learning can be used to improve the efficiency of the algorithm selection and hyperparameter optimization steps of Automated Machine Learning frameworks. In order to apply meta-learning to the LensKit-Auto framework, one must first acquire a suitable meta-dataset and then use that meta-dataset to train a meta-model. This meta-model can then be integrated into LensKit-Auto to find which LensKit algorithm is promising to achieve good performance on any given input dataset.

In this chapter, we describe the adaptation and use of an existing meta-learning framework for Automated Recommender Systems to acquire a suitable meta-dataset for use in LensKit-Auto.

\subsubsection{The meta-learning framework}

Many traditional approaches to the algorithm selection problem in Recommender Systems focus on rating prediction \cite{metaLearnerPaper}, where the datasets contain information on which items a given user interacted with, as well as how that user rated these items. A recommendation algorithm is trained on this dataset to predict the rating that a user would likely give to some item that he has not yet interacted with \cite{collFiltering, matrixFactorization}. Less research has been done on ranking prediction for implicit feedback datasets \cite{metaLearnerPaper}, where datasets only include raw user-item interactions, without any ratings. A recommendation algorithm is trained to predict which items a given user might be interested in, and to generate the top-N item recommendations for that user \cite{collFilteringImplicit, BayesianImplicit}.

To address this research gap, \citeauthor{metaLearnerPaper} propose a Meta-Learning approach in which a meta-model is trained to predict the performance of different recommendation algorithms based on the characteristics of the dataset. To achieve this, they investigate which standard meta-features of Recommender Systems datasets are meaningful for ranking prediction on implicit feedback datasets, and adapt them for this purpose. They choose 72 datasets from different domains of Recommender Systems and extract these chosen meta-features from them. Then the performance of 24 recommendation algorithms -- each with two different hyperparameter configurations -- on these datasets is evaluated and used to construct a meta-dataset. This meta-dataset is then used to train different meta-models, both traditional models like Linear Regression or Random Forest, as well as Automated Machine Learning Models with different configurations. The performance of these different meta-models on the constructed meta-dataset is then evaluated and compared \cite{metaLearnerPaper}.

The Python framework \texttt{RecSys-Algorithm-Selection-Ranking-Implicit-LBR} \cite{metaLearnerRepo} created to collect this data includes functionality for pre-processing of the input datasets, extraction of meta-features, implementations of different recommendation algorithms, evaluation using different metrics, generation of meta-datasets as well as training and evaluating different meta-models. The framework also allows parallelization of the training and evaluation steps to improve efficiency.

In order to prepare the data of different datasets for the later algorithm training and evaluation, this framework contains multiple data loader classes as well as a file \texttt{recsys\_data\_set.py}, which provides different utility functions for loading, processing and managing Recommender Systems datasets. This includes loading the raw data of the datasets specified in the paper from disk, converting them to a uniform format, and saving the processed data back to disk. The processed data for each dataset now contain the same columns and use the same file type. Some cleaning and pre-processing like removing duplicate data points or normalization of user and item IDs is also performed in this step. This file also provides functionality for splitting a dataset into train, validation, and test splits, as well as implementing 5-fold cross-validation. This file also handles the extraction of a dataset's meta-data. The meta-features considered include, for example, the number of users and items, the number of interactions, and the density of the dataset.

The supported recommendation algorithms are implemented using the external modules \texttt{LensKit} \cite{ekstrand2020lenskit, lenskitDocs}, \texttt{RecPack} \cite{recpackPaper, recpackDocs}, and \texttt{RecBole} \cite{recbolePaper, recboleDocs}. These implementations are located in \texttt{run\_lenskit.py}, \texttt{run\_recbole.py}, and \texttt{run\_recpack.py}, respectively. The wrapper file \texttt{execution\_master.py} calls functions from the suitable execution file depending on the specific algorithm name given as input parameter. These execution files handle additional pre-processing of the input data to fit the format required by the specific Recommender Systems module, including renaming of data columns or converting data to different data structures. These files each contain a \texttt{fit()} method that handles the initialization of the algorithm model given a specific model configuration, as well as training that model on the given input dataset. The set of configurations for a given algorithm model is specified in the file \texttt{algorithm\_config.py}. The trained model is then serialized using the external module \texttt{pickle} and saved to disk. In the function \texttt{predict()} the trained model is loaded from the pickle file and de-serialized. The model is then used to generate predictions of the top-20 items for each user in the dataset. These predictions are stored in a \texttt{.json} file and saved to disk. Finally, in the \texttt{evaluate()} method, these predictions are loaded and evaluated on the test data using the metrics \texttt{NDCG@K}, \texttt{HR@K} and \texttt{Recall@K} for each $K \in \{1, 3, 5, 10, 20\}$. The method calculates the mean value over all users in the dataset for each of these metrics and saves the results to disk. The specific implementation of each metric is contained in the file \texttt{run\_utils.py}

The framework also supports parallelization of the entire process of training models, generating, and evaluating recommendations. The file \texttt{hpc\_execute} automatically creates and submits a Slurm job \cite{slurm} for each independent task in the execution pipeline. Running this file on a computer cluster enables these tasks to be distributed to different nodes and executed in parallel. Each job contains either the training, prediction, or evaluation step of exactly one configuration of one specific algorithm on one specific dataset, using exactly one fold of a 5-fold cross-validation. Therefore, the total number of jobs created by this process in each step (training, prediction, and evaluation) can be expressed by 
\[
\#\text{folds} \cdot \#\text{datasets} \cdot \#\text{algorithms} \cdot \#\text{configs\_per\_algorithm}
\]

Since this framework uses 5-fold cross-validation and exactly two configurations for each algorithm, the above expression simplifies to
\[
10 \cdot \#\text{datasets} \cdot \#\text{algorithms}
\]

The file \texttt{create\_meta\_dataset.py} compiles the evaluation results of all algorithms on all datasets and generates a meta-dataset for each evaluation metric. The generated file \texttt{meta\_dataset\_NDCG@20} for example contains for each dataset its meta-features, as well as the performance of all algorithms on it, measured using the NDCG@20 metric.

Finally, the file \texttt{algorithm\_selection/process\_evaluation.py} is used to train and evaluate different meta-models on a given meta-dataset. This includes traditional meta-learning algorithms like Linear Regression or k-nearest Neighbor, as well as AutoML algorithms using the external module \texttt{AutoGluon} \cite{autogluonPaper, autogluon} with different settings. The meta-model pipeline supports the objectives \emph{algorithm performance prediction}, where the meta-model is trained to predict the performance of each algorithm on a new dataset, and \emph{algorithm ranking prediction}, where the meta-model is trained to predict just the ranking of algorithms based on their performance on a new dataset.

The meta-model with the best performance on a given meta-dataset can then be saved and integrated into an AutoRecSys framework like LensKit-Auto. However, this meta-learning framework uses LensKit version 0.14.4 and LensKit-Auto has already been updated to LensKit 2025, which introduced substantial changes to the internal architecture, impacting the behavior and performance of LensKit algorithms. This motivates the need to update this framework to also support the LensKit 2025 API as well as implement other changes to adapt the meta-learning framework specifically to LensKit-Auto. Then the adapted meta-learning framework can be used to acquire a meta-dataset and train a meta-model which is suitable for integration into LensKit-Auto.

\subsubsection{Adaptation of the Meta-Learning framework}
\label{sec:meta_learner_framework}

In order to generate a suitable meta-dataset for LensKit-Auto, we update the existing meta-learning framework to support LensKit version 2025.1.1. Since most references to the LensKit API in this project are contained within a single file -- \texttt{run\_lenskit.py} --, the entire architecture of the framework remains unchanged, only the specific implementation of LensKit within this one file is completely overhauled. The only changes outside of that one file are made to \texttt{algorithm\_config.py}, since many hyperparameters of the LensKit algorithms were renamed when switching to the 2025 API. The hyperparameter \texttt{nnrbs} of the Lenskit \texttt{ItemKNN} and \texttt{UserKNN} algorithms, for example, was renamed to \texttt{max\_nbrs} to match the name of the hyperparameter \texttt{min\_nbrs}.

The details of the update process of the \texttt{run\_lenskit.py} file closely mirror the update process of the LensKit-Auto framework to the 2025 LensKit API, as detailed in Section \ref{sec:lk_update}. First of all, the entire structure of the LensKit namespaces has changed, as well as the names of many of its classes. Therefore, all LensKit import statements, as well as many references to LensKit classes within \texttt{run\_lenskit.py}, are changed accordingly. LensKit 2025 also introduces major changes to the data structures used for the input data. Previously, LensKit handled data exclusively inside \texttt{pandas} data frames. LensKit 2025 instead implements classes that abstract from these basic data structures to specifically support the needs of Recommender systems. These data structures include the classes \texttt{Dataset}, \texttt{ItemList}, and \texttt{ItemListCollection} \cite{lenskitDocs}. This necessitates the need to add conversions to and from these new structures to the meta-learning framework. LensKit already offers many functions for that purpose, including \texttt{lenskit.data.from\_interactions\_df()}, which converts a pandas data frame to a Lenskit \texttt{Dataset} object, or \texttt{lenskit.data.ItemList.to\_df()}, which converts a Lenskit \texttt{ItemList} object back to a pandas data frame. Using these conversion methods, as seen in Listing \ref{lst:lenskit-conv}, the LensKit implementation within \texttt{run\_lenskit.py} is adapted to fit these new LensKit data structures without having to change any of the data loading and pre-processing functionality of the meta-learning framework itself.

\begin{lstlisting}[language=Python, frame=single, float=t, caption={Conversion of the input data to new Lenskit data structures}, label={lst:lenskit-conv}]
# before
train = lenskit_load_transform(
    data_set_name, fold, "train"
)

# after
data = lenskit_load_transform(
    data_set_name, fold, "train"
)
train = from_interactions_df(
    data, user_col='user', item_col='item'
)
\end{lstlisting}

Another major change introduced in LensKit 2025 is the switch to recommendation pipelines, which allow flexibly wiring together different LensKit components, such as \texttt{scorer} or \texttt{ranker}. Methods like \texttt{fit()}, \texttt{recommend()}, or \texttt{predict()} are no longer called directly on a scorer component, but instead on a pipeline containing one or more components \cite{lenskitDocs}. Since in each call of our \texttt{run\_lenskit.fit()} and \texttt{run\_lenskit.predict()} methods, we only use a single scorer component to train on the input data and generate recommendations, the switch to recommendation pipelines is rather simple. We simply initialize our scorer component exactly as before, but instead of calling \texttt{fit()} directly on our scorer, we instead initialize a top-N recommendation pipeline containing only our scorer component, and run the training method on our pipeline, as seen in Listing \ref{lst:lenskit-pipeline}. After training, we serialize and save our entire pipeline instead of just the scorer component. This trained pipeline is then used in the \texttt{run\_lenskit.predict()} method to generate recommendations.

\begin{lstlisting}[language=Python, frame=single, float=t, caption={Using a Lenskit pipeline for training}, label={lst:lenskit-pipeline}]
# before
model = Recommender.adapt(
    ItemItem(**current_configuration, feedback="implicit")
)
model.fit(train)

# after
scorer = ItemItem(
    **current_configuration, feedback="implicit"
)
pipeline = topn_pipeline(
    scorer=scorer, predicts_ratings=False
)
pipeline.train(train)
\end{lstlisting}

In addition to updating the meta-learning framework to support LensKit 2025, we also add more functionality regarding algorithm configurations. Previously, each algorithm was trained using exactly two different hyperparameter configurations, which are hard-coded in the file \texttt{algorithm\_config}. We add more possible values to most hyperparameters to create a more expansive configuration space. We then add functionality to draw $N$ different configurations from the configuration space of any given algorithm pseudo-randomly, instead of just using the first $N$ configurations. We use the module \texttt{random} to produce random numbers deterministically using the name of the algorithm and a seed. Then we use these random numbers to draw configurations from the configuration space. Therefore, for the same algorithm and using the same seed, this method will produce the exact same set of configurations every time. The seed can be passed as an optional input parameter to the \texttt{algorithm\_config.retrieve\_configurations()} method. If the user does not provide a seed, then the default seed $0$ is used. We also add the option to produce more than two configurations per algorithm by adding another parameter \texttt{num\_samples}, which can be optionally passed to this method. If the user does not provide such a number, the default value $2$ is used, to keep the behavior consistent with previous versions of the framework.

Lastly, since we intend to use the meta-learning framework to produce a meta-dataset that is suitable specifically for LensKit-Auto, we decide to use the evaluation metric implementations provided by LensKit instead of the manual implementations used previously. Even slight differences in the implementations can lead to a different ordering of algorithms according to their performance. By using the same metric implementations in both the meta-learning framework and in LensKit-Auto, we ensure that these differences do not negatively impact the accuracy of the meta-model when integrated into LensKit-Auto. We add a method \texttt{metrics\_lenskit()} to the file \texttt{run\_utils.py}, which initializes a \texttt{lenskit.metrics.RunAnalysis} object with all desired metrics and computes the evaluation results using \texttt{RunAnalysis.measure()}. We change each execution file to use \texttt{metrics\_lenskit()} when evaluating recommendations.

\subsubsection{Experimental Execution and Meta-Dataset Generation}

To generate a meta-dataset suitable for LensKit-Auto, we execute the full training, prediction, and evaluation pipeline of the adapted meta-learning framework across all combinations of available datasets and available algorithms, using five hyperparameter configurations per algorithm. Each individual experiment corresponds to either the training, prediction, or evaluation step of a single algorithm with a specific configuration on a single dataset and one fold of a 5-fold cross-validation. Using the formula given in Section \ref{sec:meta_learner_framework} for 23 algorithms and 66 datasets, the number of jobs created per pipeline step is equal to $23 \cdot 66 \cdot 5 \cdot 5 = 37950$, or $37950 \cdot 3 = 113850$ jobs in total.

We run these experiments on the OMNI cluster of the University of Siegen (AMD EPYC 7452, Tesla V100) \cite{cluster} using a Slurm-based job scheduling system. As detailed in Section \ref{sec:meta_learner_framework}, the meta-learning framework includes functionality to automatically create a separate Slurm job for each independent unit of work in the pipeline. This allows experiments to be executed in parallel across multiple compute nodes. This parallel execution strategy makes it feasible to generate the entire meta-dataset within a reasonable time frame. The specific command run on a login node of the cluster to produce and submit all jobs for the training step of the pipeline can be seen in Listing \ref{lst:training_cmd}. The results of all evaluation jobs are collected and aggregated using the \texttt{create\_meta\_dataset.py} script, producing one meta-dataset per evaluation metric. Each data point of one of these meta-datasets corresponds to a dataset described by its meta-features, along with the performance of all evaluated algorithms on that dataset, measured by a specific metric.

\begin{lstlisting}[language=bash, caption={The command run on the cluster to produce jobs for the training step}, label={lst:training_cmd}, frame=single, float=t, basicstyle=\ttfamily\small]
python hpc_execute.py --mode fit --num_samples 5 --seed 0
\end{lstlisting}

The generated meta-datasets can be found in the \texttt{algorithm\_selection} folder of the adapted meta-learning framework \cite{metaLearnerRepoNew}. These results can form the basis for a future analysis of the potential benefit of integrating meta-learning into the LensKit-Auto framework more generally, as well as providing the possibility to train a specific meta-model for subsequent integration.

\section{Conclusion}
This report summarizes the work carried out as part of the project group that worked on updating and improving the LensKit-Auto Library. Recommender systems require systematic evaluation and careful configuration of hyperparameter, as the model performance depends strongly on both algorithm choice and parameter settings \cite{Aggarwal2016Recommender, Bergstra2011}

The project group first focused on updating LensKit-Auto to restore compatibility with recent versions of the LensKit framework and to improve the usability and robustness of the existing codebase. This was achieved by adapting the framework to the updated LensKit 2025 API and revising components that were no longer compatible with the new data structures and interfaces. 

Focus was also placed on improving the testing infrastructure. All tests were revised to better reflect the changes introduced in the library. The updated tests are able to identify inconsistencies in data handling and parameter processing, making it easier to identify and fix breaking changes. The documentation was updated to reflect the current API and to provide clearer usage guidance.

Model persistence functionality was introduced to support reuse of trained models and configurations across experiments. This feature enables the storage of models together with their corresponding hyperparameter configurations, allowing experiments to be reproduced without retraining. It simplifies the workflow for evaluating previously identified configurations and reduces redundant computations during iterative experiments. 

In addition, integration with the DeepCave framework enables analysis of the hyperparameter optimization process. DeepCave provides visualization tools for exploring the search space and understanding the relationship between hyperparameter configurations and performance. This allows users to inspect optimization trajectories, compare configurations, and identify regions of the search space associated with strong performance. 

A central contribution of this work is the integration of the Tree Parzen Estimator (\textbf{TPE}) as an additional optimization strategy. TPE extends the set of available optimization methods by providing a density-based Bayesian optimization approach that can handle structured and conditional hyperparameter spaces efficiently \cite{Bergstra2011}. This implementation seamlessly integrates into the existing optimization pipeline and allows users to select TPE alongside the existing strategies Random Search and Bayesian optimization.

Finally, we adapted an existing meta-learning framework for Recommender Systems to ensure compatibility with the updated LensKit 2025 API and the LensKit-Auto framework. We extended the configuration spaces of the framework and changed its evaluation pipeline to use LensKit's metric implementations to align it with the evaluation process of LensKit-Auto. Using the adapted framework, we conducted large-scale experiments on a computing cluster to generate meta-datasets that capture the performance of various recommendation algorithms across multiple datasets. These meta-datasets provide the basis for training meta-models and enable future integration of meta-learning-based algorithm selection into LensKit-Auto.

The resulting system provides a more robust and extensible platform for the experimentation of automated recommender systems. By combining updated framework compatibility, improved testing and documentation, additional optimization capabilities, and visualization support, LensKit-Auto is better suited for structured and reproducible experimentation workflows. 

Future work may focus on extending optimization capabilities further, improving scalability, and integrating additional techniques such as meta-learning to guide algorithm selection and configuration more efficiently. 

\section{Acknowledgments}
We acknowledge the use of ChatGPT to refine the clarity and flow of our writing, and DeepL to assist with accurate translation. Both tools were employed solely as aids for grammar, phrasing, and formulation.

\newpage
\listoffigures
\listoftables
\lstlistoflistings
\newpage

\bibliography{bibliography}{}

@inproceedings{datasetCharecteristics,
author = {Wagne, Ahmadou and Neidhardt, Julia},
title = {What to compare? Towards understanding user sessions on price comparison platforms},
year = {2024},
isbn = {9798400705052},
publisher = {Association for Computing Machinery},
address = {New York, NY, USA},
url = {https://doi.org/10.1145/3640457.3691717},
doi = {10.1145/3640457.3691717},
booktitle = {Proceedings of the 18th ACM Conference on Recommender Systems},
pages = {1158–1162},
numpages = {5},
location = {Bari, Italy},
series = {RecSys '24}
}

@inproceedings{10.1145/3604915.3609488,
author = {Shehzad, Faisal and Jannach, Dietmar},
title = {Everyone’s a Winner! On Hyperparameter Tuning of Recommendation Models},
year = {2023},
isbn = {9798400702419},
publisher = {Association for Computing Machinery},
address = {New York, NY, USA},
url = {https://doi.org/10.1145/3604915.3609488},
doi = {10.1145/3604915.3609488},
booktitle = {Proceedings of the 17th ACM Conference on Recommender Systems},
pages = {652–657},
numpages = {6},
location = {Singapore, Singapore},
series = {RecSys '23}
}

@article{EvaluationProtocols,
author = {Cremonesi, Paolo and Jannach, Dietmar},
title = {Progress in recommender systems research: Crisis? What crisis?},
journal = {AI Magazine},
volume = {42},
number = {3},
pages = {43-54},
doi = {https://doi.org/10.1609/aimag.v42i3.18145},
url = {https://onlinelibrary.wiley.com/doi/abs/10.1609/aimag.v42i3.18145},
eprint = {https://onlinelibrary.wiley.com/doi/pdf/10.1609/aimag.v42i3.18145},
year = {2021}
}

@article{HyperparameterSearch,
  author  = {James Bergstra and Yoshua Bengio},
  title   = {Random Search for Hyper-Parameter Optimization},
  journal = {Journal of Machine Learning Research},
  year    = {2012},
  volume  = {13},
  number  = {10},
  pages   = {281--305},
  url     = {http://jmlr.org/papers/v13/bergstra12a.html}
}

@book{AutoMLPopularity,
  author    = {Frank Hutter and Lars Kotthoff and Joaquin Vanschoren},
  title     = {Automated Machine Learning: Methods, Systems, Challenges},
  publisher = {Springer},
  year      = {2019},
  doi       = {10.1007/978-3-030-05318-5}
}

@ARTICLE{PersonalizedRecs,
  author={Adomavicius, G. and Tuzhilin, A.},
  journal={IEEE Transactions on Knowledge and Data Engineering}, 
  title={Toward the next generation of recommender systems: a survey of the state-of-the-art and possible extensions}, 
  year={2005},
  volume={17},
  number={6},
  pages={734-749},
  doi={10.1109/TKDE.2005.99}}

@article{Parzen1962,
 ISSN = {00034851},
 URL = {http://www.jstor.org/stable/2237880},
 author = {Emanuel Parzen},
 journal = {The Annals of Mathematical Statistics},
 number = {3},
 pages = {1065--1076},
 publisher = {Institute of Mathematical Statistics},
 title = {On Estimation of a Probability Density Function and Mode},
 urldate = {2026-03-31},
 volume = {33},
 year = {1962}
}

@book{Discovery,
  editor    = {Francesco Ricci and Lior Rokach and Bracha Shapira and Paul B. Kantor},
  title     = {Recommender Systems Handbook},
  publisher = {Springer},
  year      = {2011},
  doi       = {10.1007/978-0-387-85820-3}
}

@inbook{WideUse,
author = {Amatriain, Xavier and Basilico, Justin},
year = {2015},
month = {01},
pages = {385-419},
title = {Recommender Systems in Industry: A Netflix Case Study},
isbn = {978-1-4899-7636-9},
doi = {10.1007/978-1-4899-7637-6_11}
}

@book{Aggarwal2016Recommender,
  title={Recommender Systems: The Textbook},
  author={Aggarwal, Charu C.},
  publisher={Springer},
  year={2016},
  doi = {10.1007/978-3-319-29659-3}
}

@article{Su2009CollaborativeFiltering,
  author = {Su, Xiaoyuan and Khoshgoftaar, Taghi M.},
  title = {A Survey of Collaborative Filtering Techniques},
  journal = {Advances in Artificial Intelligence},
  volume = {2009},
  number = {1},
  pages = {421425},
  doi = {https://doi.org/10.1155/2009/421425},
  url = {https://onlinelibrary.wiley.com/doi/abs/10.1155/2009/421425},
}

@inproceedings{Sarwar2001ItemBasedCF,
    author = {Sarwar, Badrul and Karypis, George and Konstan, Joseph and Riedl, John},
    title = {Item-based collaborative filtering recommendation algorithms},
    year = {2001},
    isbn = {1581133480},
    publisher = {Association for Computing Machinery},
    address = {New York, NY, USA},
    url = {https://doi.org/10.1145/371920.372071},
    doi = {10.1145/371920.372071},
    booktitle = {Proceedings of the 10th International Conference on World Wide Web},
    pages = {285–295},
    numpages = {11},
    location = {Hong Kong, Hong Kong},
    series = {WWW '01}
}

@InProceedings{meta_learning_blackbox,
  title = 	 {Opening the Black Box: Automated Software Analysis for Algorithm Selection},
  author =       {Pulatov, Damir and Anastacio, Marie and Kotthoff, Lars and Hoos, Holger},
  booktitle = 	 {Proceedings of the First International Conference on Automated Machine Learning},
  pages = 	 {6/1--18},
  year = 	 {2022},
  editor = 	 {Guyon, Isabelle and Lindauer, Marius and van der Schaar, Mihaela and Hutter, Frank and Garnett, Roman},
  volume = 	 {188},
  series = 	 {Proceedings of Machine Learning Research},
  month = 	 {25--27 Jul},
  publisher =    {PMLR},
  url = 	 {https://proceedings.mlr.press/v188/pulatov22a.html}
}

@article{Koren2009MatrixFactorization,
  author={Koren, Yehuda and Bell, Robert and Volinsky, Chris},
  journal={Computer}, 
  title={Matrix Factorization Techniques for Recommender Systems}, 
  year={2009},
  volume={42},
  number={8},
  pages={30-37},
  doi={10.1109/MC.2009.263}
}

@INPROCEEDINGS{Hu2008Implicit,
  author={Hu, Yifan and Koren, Yehuda and Volinsky, Chris},
  booktitle={2008 Eighth IEEE International Conference on Data Mining}, 
  title={Collaborative Filtering for Implicit Feedback Datasets}, 
  year={2008},
  volume={},
  number={},
  pages={263-272},
  doi={10.1109/ICDM.2008.22}
}

@inproceedings{Bergstra2011,
    author = {Bergstra, James and Bardenet, R\'{e}mi and Bengio, Yoshua and K\'{e}gl, Bal\'{a}zs},
    title = {Algorithms for hyper-parameter optimization},
    year = {2011},
    isbn = {9781618395993},
    publisher = {Curran Associates Inc.},
    address = {Red Hook, NY, USA},
    booktitle = {Proceedings of the 25th International Conference on Neural Information Processing Systems},
    pages = {2546–2554},
    numpages = {9},
    location = {Granada, Spain},
    series = {NIPS'11}
}

@book{Feurer2019AutoML,
    author = {Hutter, Frank and Kotthoff, Lars and Vanschoren, Joaquin},
    year = {2019},
    month = {01},
    pages = {},
    title = {Automated Machine Learning - Methods, Systems, Challenges},
    isbn = {978-3-030-05317-8},
    doi = {10.1007/978-3-030-05318-5}
}

@inproceedings{Hutter2011SMAC,
    author = {Hutter, Frank and Hoos, Holger H. and Leyton-Brown, Kevin},
    title = {Sequential model-based optimization for general algorithm configuration},
    year = {2011},
    isbn = {9783642255656},
    publisher = {Springer-Verlag},
    address = {Berlin, Heidelberg},
    url = {https://doi.org/10.1007/978-3-642-25566-3_40},
    doi = {10.1007/978-3-642-25566-3_40},
    booktitle = {Proceedings of the 5th International Conference on Learning and Intelligent Optimization},
    pages = {507–523},
    numpages = {17},
    location = {Rome, Italy},
    series = {LION'05}
}

@misc{Snoek2012BO,
      title={Practical Bayesian Optimization of Machine Learning Algorithms}, 
      author={Jasper Snoek and Hugo Larochelle and Ryan P. Adams},
      year={2012},
      eprint={1206.2944},
      archivePrefix={arXiv},
      primaryClass={stat.ML},
      url={https://arxiv.org/abs/1206.2944}, 
}

@article{Shahriari2016BO,
    title = "Taking the human out of the loop: A review of Bayesian optimization",
    author = "Bobak Shahriari and Kevin Swersky and Ziyu Wang and Adams, \{Ryan P.\} and \{De Freitas\}, Nando",
    note = "Publisher Copyright: {\textcopyright} 1963-2012 IEEE.",
    year = "2016",
    month = jan,
    doi = "10.1109/JPROC.2015.2494218",
    language = "English (US)",
    volume = "104",
    pages = "148--175",
    journal = "Proceedings of the IEEE",
    issn = "0018-9219",
    publisher = "Institute of Electrical and Electronics Engineers Inc.",
    number = "1",
}

@inproceedings{Feurer2015AutoSklearn,
 author = {Feurer, Matthias and Klein, Aaron and Eggensperger, Katharina and Springenberg, Jost and Blum, Manuel and Hutter, Frank},
 booktitle = {Advances in Neural Information Processing Systems},
 editor = {C. Cortes and N. Lawrence and D. Lee and M. Sugiyama and R. Garnett},
 pages = {},
 publisher = {Curran Associates, Inc.},
 title = {Efficient and Robust Automated Machine Learning},
 url = {https://proceedings.neurips.cc/paper_files/paper/2015/file/11d0e6287202fced83f79975ec59a3a6-Paper.pdf},
 volume = {28},
 year = {2015}
}

@article{koren2009matrix,
  title={Matrix factorization techniques for recommender systems},
  author={Koren, Yehuda and Bell, Robert and Volinsky, Chris},
  journal={Computer},
  volume={42},
  number={8},
  pages={30--37},
  year={2009},
  publisher={IEEE}
}

@inproceedings{hu2008collaborative,
  title={Collaborative filtering for implicit feedback datasets},
  author={Hu, Yifan and Koren, Yehuda and Volinsky, Chris},
  booktitle={2008 Eighth IEEE international conference on data mining},
  pages={263--272},
  year={2008},
  organization={Ieee}
}

@incollection{gunawardana2012evaluating,
  title={Evaluating recommender systems},
  author={Gunawardana, Asela and Shani, Guy and Yogev, Sivan},
  booktitle={Recommender systems handbook},
  pages={547--601},
  year={2012},
  publisher={Springer}
}

@article{jarvelin2002cumulated,
  title={Cumulated gain-based evaluation of IR techniques},
  author={J{\"a}rvelin, Kalervo and Kek{\"a}l{\"a}inen, Jaana},
  journal={ACM Transactions on Information Systems (TOIS)},
  volume={20},
  number={4},
  pages={422--446},
  year={2002},
  publisher={ACM New York, NY, USA}
}

@article{herlocker2004evaluating,
  title={Evaluating collaborative filtering recommender systems},
  author={Herlocker, Jonathan L and Konstan, Joseph A and Terveen, Loren G and Riedl, John T},
  journal={ACM Transactions on Information Systems (TOIS)},
  volume={22},
  number={1},
  pages={5--53},
  year={2004},
  publisher={ACM New York, NY, USA}
}

@article{deshpande2004item,
  title={Item-based top-n recommendation algorithms},
  author={Deshpande, Mukund and Karypis, George},
  journal={ACM Transactions on Information Systems (TOIS)},
  volume={22},
  number={1},
  pages={143--177},
  year={2004},
  publisher={ACM New York, NY, USA}
}

@article{gunawardana2009survey,
  title={A survey of accuracy evaluation metrics of recommendation tasks.},
  author={Gunawardana, Asela and Shani, Guy},
  journal={Journal of Machine Learning Research},
  volume={10},
  number={12},
  year={2009}
}

@incollection{trattner2023evaluating,
  title={Evaluating group recommender systems},
  author={Trattner, Christoph and Said, Alan and Boratto, Ludovico and Felfernig, Alexander},
  booktitle={Group recommender systems: an introduction},
  pages={63--75},
  year={2023},
  publisher={Springer}
}

@inproceedings{thornton2013auto,
  title={Auto-WEKA: Combined selection and hyperparameter optimization of classification algorithms},
  author={Thornton, Chris and Hutter, Frank and Hoos, Holger H and Leyton-Brown, Kevin},
  booktitle={Proceedings of the 19th ACM SIGKDD international conference on Knowledge discovery and data mining},
  pages={847--855},
  year={2013}
}

@inproceedings{vente2023introducing,
  title={Introducing lenskit-auto, an experimental automated recommender system (autorecsys) toolkit},
  author={Vente, Tobias and Ekstrand, Michael and Beel, Joeran},
  booktitle={Proceedings of the 17th ACM Conference on Recommender Systems},
  pages={1212--1216},
  year={2023}
}

@book{hutter2019automated,
  title={Automated machine learning: methods, systems, challenges},
  author={Hutter, Frank and Kotthoff, Lars and Vanschoren, Joaquin},
  year={2019},
  publisher={Springer}
}

@inproceedings{anand2020auto,
  title={Auto-surprise: An automated recommender-system (autorecsys) library with tree of parzens estimator (tpe) optimization},
  author={Anand, Rohan and Beel, Joeran},
  booktitle={Proceedings of the 14th ACM Conference on Recommender Systems},
  pages={585--587},
  year={2020}
}

@article{zheng2023automl,
  title={Automl for deep recommender systems: A survey},
  author={Zheng, Ruiqi and Qu, Liang and Cui, Bin and Shi, Yuhui and Yin, Hongzhi},
  journal={ACM Transactions on Information Systems},
  volume={41},
  number={4},
  pages={1--38},
  year={2023},
  publisher={ACM New York, NY}
}

@misc{deepcave,
      title={DeepCAVE: An Interactive Analysis Tool for Automated Machine Learning}, 
      author={René Sass and Eddie Bergman and André Biedenkapp and Frank Hutter and Marius Lindauer},
      year={2022},
      eprint={2206.03493},
      archivePrefix={arXiv},
      primaryClass={cs.LG},
      url={https://arxiv.org/abs/2206.03493}, 
}

@inproceedings{ekstrand2020lenskit,
  title={Lenskit for python: Next-generation software for recommender systems experiments},
  author={Ekstrand, Michael D},
  booktitle={Proceedings of the 29th ACM international conference on information \& knowledge management},
  pages={2999--3006},
  year={2020}
}

@inproceedings{meta_learning,
    author = {Vanschoren, Joaquin},
    title = {Meta-Learning},
    pages = {39-68},
    chapter = {2}
}

@article{meta_learning_recsys,
title = {Metalearning and Recommender Systems: A literature review and empirical study on the algorithm selection problem for Collaborative Filtering},
journal = {Information Sciences},
volume = {423},
pages = {128-144},
year = {2018},
issn = {0020-0255},
doi = {https://doi.org/10.1016/j.ins.2017.09.050},
url = {https://www.sciencedirect.com/science/article/pii/S0020025517309702},
author = {Tiago Cunha and Carlos Soares and André C.P.L.F. {de Carvalho}},
}

@misc{metaLearnerRepo,
title={Meta-learning framework Git Repository},
howpublished={\url{https://github.com/ISG-Siegen/RecSys-Algorithm-Selection-Ranking-Implicit-LBR}},
note={Accessed: 31.03.2026}
}

@misc{metaLearnerRepoNew,
title={Git Repository of the adapted Meta-learning framework},
howpublished={\url{https://github.com/LucaQuade/RecSys-Algorithm-Selection-Ranking-Implicit-LBR}},
note={Accessed: 31.03.2026}
}

@misc{movieLens,
title={MovieLens datasets},
howpublished={\url{https://cseweb.ucsd.edu/~jmcauley/datasets/amazon/links.html}},
note={Accessed: 31.03.2026}
}

@misc{amazon,
title={Amazon2014 datasets},
howpublished={\url{https://grouplens.org/datasets/movielens/}},
note={Accessed: 31.03.2026}
}

@article{collFiltering,
author = {Ekstrand, Michael D. and Riedl, John T. and Konstan, Joseph A.},
title = {Collaborative Filtering Recommender Systems},
year = {2011},
issue_date = {February 2011},
publisher = {Now Publishers Inc.},
address = {Hanover, MA, USA},
volume = {4},
number = {2},
issn = {1551-3955},
url = {https://doi.org/10.1561/1100000009},
doi = {10.1561/1100000009},
journal = {Found. Trends Hum.-Comput. Interact.},
month = feb,
pages = {81–173},
numpages = {93}
}

@inproceedings{collFilteringImplicit,
author = {Hu, Yifan and Koren, Yehuda and Volinsky, Chris},
title = {Collaborative Filtering for Implicit Feedback Datasets},
year = {2008},
isbn = {9780769535029},
publisher = {IEEE Computer Society},
address = {USA},
url = {https://doi.org/10.1109/ICDM.2008.22},
doi = {10.1109/ICDM.2008.22},
booktitle = {Proceedings of the 2008 Eighth IEEE International Conference on Data Mining},
pages = {263–272},
numpages = {10},
series = {ICDM '08}
}

@inproceedings{BayesianImplicit,
author = {Rendle, Steffen and Freudenthaler, Christoph and Gantner, Zeno and Schmidt-Thieme, Lars},
title = {BPR: Bayesian personalized ranking from implicit feedback},
year = {2009},
isbn = {9780974903958},
publisher = {AUAI Press},
address = {Arlington, Virginia, USA},
booktitle = {Proceedings of the Twenty-Fifth Conference on Uncertainty in Artificial Intelligence},
pages = {452–461},
numpages = {10},
location = {Montreal, Quebec, Canada},
series = {UAI '09}
}

@article{matrixFactorization,
author = {Koren, Yehuda and Bell, Robert and Volinsky, Chris},
title = {Matrix Factorization Techniques for Recommender Systems},
year = {2009},
issue_date = {August 2009},
publisher = {IEEE Computer Society Press},
address = {Washington, DC, USA},
volume = {42},
number = {8},
issn = {0018-9162},
url = {https://doi.org/10.1109/MC.2009.263},
doi = {10.1109/MC.2009.263},
journal = {Computer},
month = aug,
pages = {30–37},
numpages = {8}
}

@inproceedings{metaLearnerPaper,
author = {Wegmeth, Lukas and Vente, Tobias and Beel, Joeran},
title = {Recommender Systems Algorithm Selection for Ranking Prediction on Implicit Feedback Datasets},
year = {2024},
isbn = {9798400705052},
publisher = {Association for Computing Machinery},
address = {New York, NY, USA},
url = {https://doi.org/10.1145/3640457.3691718},
doi = {10.1145/3640457.3691718},
booktitle = {Proceedings of the 18th ACM Conference on Recommender Systems},
pages = {1163–1167},
numpages = {5},
location = {Bari, Italy},
series = {RecSys '24}
}

@misc{deepcaveDocs,
title={DeepCAVE documentation},
howpublished={\url{https://automl.github.io/DeepCAVE/main/index.html}},
note={Accessed: 30.03.2026}
}

@inproceedings{arabzadeh2024green,
  title={Green recommender systems: Optimizing dataset size for energy-efficient algorithm performance},
  author={Arabzadeh, Ardalan and Vente, Tobias and Beel, Joeran},
  booktitle={International Workshop on Recommender Systems for Sustainability and Social Good},
  pages={73--82},
  year={2024},
  organization={Springer}
}

@inproceedings{vente2024greedy,
  title={Greedy Ensemble Selection for Top-N Recommendations.},
  author={Vente, Tobias and Mehta, Zainil and Wegmeth, Lukas and Beel, Joeran},
  booktitle={RobustRecSys@ RecSys},
  pages={12--16},
  year={2024}
}

@misc{lenskitDocs,
title={LensKit documentation},
howpublished={\url{https://lenskit.org/stable/index.html}},
note={Accessed: 31.03.2026}
}

@misc{recboleDocs,
title={RecBole documentation},
howpublished={\url{https://recbole.io/docs/}},
note={Accessed: 31.03.2026}
}

@inproceedings{recbolePaper,
author = {Zhao, Wayne Xin and Mu, Shanlei and Hou, Yupeng and Lin, Zihan and Chen, Yushuo and Pan, Xingyu and Li, Kaiyuan and Lu, Yujie and Wang, Hui and Tian, Changxin and Min, Yingqian and Feng, Zhichao and Fan, Xinyan and Chen, Xu and Wang, Pengfei and Ji, Wendi and Li, Yaliang and Wang, Xiaoling and Wen, Ji-Rong},
title = {RecBole: Towards a Unified, Comprehensive and Efficient Framework for Recommendation Algorithms},
year = {2021},
isbn = {9781450384469},
publisher = {Association for Computing Machinery},
address = {New York, NY, USA},
url = {https://doi.org/10.1145/3459637.3482016},
doi = {10.1145/3459637.3482016},
booktitle = {Proceedings of the 30th ACM International Conference on Information \& Knowledge Management},
pages = {4653–4664},
numpages = {12},
location = {Virtual Event, Queensland, Australia},
series = {CIKM '21}
}

@misc{recpackDocs,
title={RecPack documentation},
howpublished={\url{https://recpack.froomle.ai/}},
note={Accessed: 31.03.2026}
}

@inproceedings{recpackPaper,
author = {Michiels, Lien and Verachtert, Robin and Goethals, Bart},
title = {RecPack: An(other) Experimentation Toolkit for Top-N Recommendation using Implicit Feedback Data},
year = {2022},
isbn = {9781450392785},
publisher = {Association for Computing Machinery},
address = {New York, NY, USA},
url = {https://doi.org/10.1145/3523227.3551472},
doi = {10.1145/3523227.3551472},
booktitle = {Proceedings of the 16th ACM Conference on Recommender Systems},
pages = {648–651},
numpages = {4},
location = {Seattle, WA, USA},
series = {RecSys '22}
}

@misc{slurm,
title={Slurm documentation},
howpublished={\url{https://slurm.schedmd.com/overview.html}},
note={Accessed: 31.03.2026}
}

@misc{autogluon,
title={AutoGluon documentation},
howpublished={\url{https://auto.gluon.ai/stable/index.html}},
note={Accessed: 31.03.2026}
}

@misc{cluster,
title={Homepage of the OMNI cluster of the University of Siegen},
howpublished={\url{https://cluster.uni-siegen.de/}},
note={Accessed: 31.03.2026}
}

@article{autogluonPaper,
  title={AutoGluon-Tabular: Robust and Accurate AutoML for Structured Data},
  author={Nick Erickson and Jonas W. Mueller and Alexander Shirkov and Hang Zhang and Pedro Larroy and Mu Li and Alex Smola},
  journal={ArXiv},
  year={2020},
  volume={abs/2003.06505},
  url={https://api.semanticscholar.org/CorpusID:212725762}
}

@article{bergstra2012random,
  title={Random search for hyper-parameter optimization.},
  author={Bergstra, James and Bengio, Yoshua},
  journal={Journal of machine learning research},
  volume={13},
  number={2},
  year={2012}
}

@article{snoek2012practical,
  title={Practical bayesian optimization of machine learning algorithms},
  author={Snoek, Jasper and Larochelle, Hugo and Adams, Ryan P},
  journal={Advances in neural information processing systems},
  volume={25},
  year={2012}
}

@inproceedings{vente2023advancing,
  title={Advancing automation of design decisions in recommender system pipelines},
  author={Vente, Tobias},
  booktitle={Proceedings of the 17th ACM Conference on Recommender Systems},
  pages={1355--1360},
  year={2023}
}

@inproceedings{AutoSurprise,
author = {Anand, Rohan and Beel, Joeran},
title = {Auto-Surprise: An Automated Recommender-System (AutoRecSys) Library with Tree of Parzens Estimator (TPE) Optimization},
year = {2020},
isbn = {9781450375832},
publisher = {Association for Computing Machinery},
address = {New York, NY, USA},
url = {https://doi.org/10.1145/3383313.3411467},
doi = {10.1145/3383313.3411467},
booktitle = {Proceedings of the 14th ACM Conference on Recommender Systems},
pages = {585–587},
numpages = {3},
location = {Virtual Event, Brazil},
series = {RecSys '20}
}

@inproceedings{librecAuto,
author = {Sonboli, Nasim and Mansoury, Masoud and Guo, Ziyue and Kadekodi, Shreyas and Liu, Weiwen and Liu, Zijun and Schwartz, Andrew and Burke, Robin},
title = {librec-auto: A Tool for Recommender Systems Experimentation},
year = {2021},
isbn = {9781450384469},
publisher = {Association for Computing Machinery},
address = {New York, NY, USA},
url = {https://doi.org/10.1145/3459637.3482006},
doi = {10.1145/3459637.3482006},
booktitle = {Proceedings of the 30th ACM International Conference on Information \& Knowledge Management},
pages = {4584–4593},
numpages = {10},
location = {Virtual Event, Queensland, Australia},
series = {CIKM '21}
}

@inproceedings{vente2025potential,
  title={The potential of automl for recommender systems},
  author={Vente, Tobias and Wegmeth, Lukas and Beel, Joeran},
  booktitle={Adjunct Proceedings of the 33rd ACM Conference on User Modeling, Adaptation and Personalization},
  pages={371--378},
  year={2025}
}

@inproceedings{vente2024clicks,
  title={From clicks to carbon: the environmental toll of recommender systems},
  author={Vente, Tobias and Wegmeth, Lukas and Said, Alan and Beel, Joeran},
  booktitle={Proceedings of the 18th ACM Conference on Recommender Systems},
  pages={580--590},
  year={2024}
}

@article{wegmeth2025green,
  title={Green Recommender Systems: Understanding and Minimizing the Carbon Footprint of AI-Powered Personalization},
  author={Wegmeth, Lukas and Vente, Tobias and Said, Alan and Beel, Joeran},
  journal={ACM Transactions on Recommender Systems},
  year={2025},
  publisher={ACM New York, NY}
}

@inproceedings{vente2025aps,
  title={APS Explorer: Navigating Algorithm Performance Spaces for Informed Dataset Selection},
  author={Vente, Tobias and Heep, Michael and Abbas, Abdullah and Sperle, Theodor and Beel, Joeran and Goethals, Bart},
  booktitle={Proceedings of the Nineteenth ACM Conference on Recommender Systems},
  pages={1322--1324},
  year={2025}
}

@inproceedings{wegmeth2024recommender,
  title={Recommender systems algorithm selection for ranking prediction on implicit feedback datasets},
  author={Wegmeth, Lukas and Vente, Tobias and Beel, Joeran},
  booktitle={Proceedings of the 18th ACM Conference on Recommender Systems},
  pages={1163--1167},
  year={2024}
}

@article{beel20224,
  title={4.3 Best-Practices for Offline Evaluations of Recommender Systems},
  author={Beel, Joeran and Jannach, Dietmar and Said, Alan and Shani, Guy and Vente, Tobias and Wegmeth, Lukas},
  journal={Evaluation Perspectives of Recommender Systems: Driving Research and Education},
  volume={55},
  number={8},
  pages={110},
  year={2022}
}
\bibliographystyle{plainnat}

\end{document}